\documentclass[lettersize,journal]{IEEEtran}
\usepackage{amsmath,amsfonts}
\usepackage{algorithmic}
\usepackage{algorithm}
\usepackage{array}
\usepackage{textcomp}
\usepackage{stfloats}
\usepackage{url}
\usepackage{verbatim}
\usepackage{graphicx}
\usepackage{cite}

\usepackage{listings} 
\usepackage{gensymb}
\usepackage{graphicx}
\usepackage{mathtools}
\usepackage[inline]{enumitem}
\usepackage{amsmath}
\usepackage{amssymb}
\usepackage{optidef}
\usepackage{lineno,hyperref}
\usepackage{amsthm}
\usepackage{xcolor}
\usepackage{algorithm,algorithmic}
\usepackage{subcaption}
\usepackage{bbm}
\usepackage{eqparbox}
\usepackage{makecell}
\usepackage{booktabs}

\usepackage{etoolbox}
\usepackage{calc}
\usepackage{upgreek}
\usepackage{multirow}
\usepackage{adjustbox}
\usepackage{lscape}
\usepackage{pdfpages}
\usepackage{subcaption}
\usepackage{wrapfig}
\usepackage{pifont} 
\newcommand{\yesmark}{\ding{51}}  
\newcommand{\nomark}{\ding{55}}
\begin{document}

\title{SOUL: A \underline{S}emi-supervised \underline{O}pen-world contin\underline{U}al \underline{L}earning method for Network Intrusion Detection}

\author{Suresh Kumar Amalapuram, Shreya Kumar, Bheemarjuna Reddy Tamma,~\IEEEmembership{Senior Member,~IEEE,} \\Sumohana Channappayya,~\IEEEmembership{Senior Member,~IEEE,}
}



\maketitle

\begin{abstract}
Fully supervised continual learning methods have shown improved attack traffic detection in a closed-world learning setting. However, obtaining fully annotated data is an arduous task in the security domain. Further, our research finds that after training a classifier on two days of network traffic, the performance decay of attack class detection over time (computed using the area under the time on precision-recall AUC of the attack class) drops from 0.985 to 0.506 on testing with three days of new test samples. In this work, we focus on label scarcity and open-world learning (OWL) settings to improve the attack class detection of the continual learning-based network intrusion detection (NID). We formulate OWL for NID as a semi-supervised continual learning-based method, dubbed SOUL, to achieve the classifier performance on par with fully supervised models while using limited annotated data. The proposed method is motivated by our empirical observation that using gradient projection memory (constructed using buffer memory samples) can significantly improve the detection performance of the attack (minority) class when trained using partially labeled data. Further, using the classifier's confidence in conjunction with buffer memory, SOUL generates high-confidence labels whenever it encounters OWL tasks closer to seen tasks, thus acting as a label generator. Interestingly, SOUL efficiently utilizes samples in the buffer memory for sample replay to avoid catastrophic forgetting, construct the projection memory, and assist in generating labels for unseen tasks. The proposed method is evaluated on four standard network intrusion detection datasets, and the performance results are closer to the fully supervised baselines using at most 20\% labeled data while reducing the data annotation effort in the range of 11 to 45\% for unseen data.
\end{abstract}


\section{Introduction}
In its cybersecurity readiness index~\cite{ciscomar2023}, Cisco categorizes the organization's preparedness for cyber threats and identifies that securing the network from malicious actors is critical to the viability of an organization. Network Intrusion Detection Systems (NIDS)~\cite{smaha1988haystack,10.1145/1541880.1541882} are an important cybersecurity system/tool to secure a network from cyber threats. Existing NIDS must evolve continuously to combat the ever-evolving cyber attacks~\cite{chow2023drift,andresini2021insomnia,amalapuram2023augmented}. The continuous evolution of such systems requires to meet two desiderata \begin{enumerate*}[label=(\roman*)]
    \item the ability to detect known threats and
    \item leverage the learned knowledge to identify and adapt to novel/unseen attacks
     
\end{enumerate*}. Specifically, existing systems fail to satisfy the first desideratum, which is said to exhibit \textit{catastrophic forgetting} (CF)~\cite{mai2022online} of the learned knowledge. The second desideratum is also known as \textit{open-world learning}. 

\par To reduce the effects of CF, the paradigm of continual learning (CL)~\cite{thrun1995lifelong,parisi2019continual} can be employed. The direct application of CL to security applications is under-explored. Existing CL literature~\cite{chaudhry2018efficient,aljundi2019online,parisi2019continual,aljundi2019gradient,de2021continual,wang2023comprehensive} focuses on supervised learning settings requiring annotated data. However, obtaining labeled data is difficult in cybersecurity applications. The complexity of the data annotation in cybersecurity comes from different facets: sample labelling complexity, labelling budget~\cite{apruzzese2022sok}, and implementation gap~\cite{apruzzese2023role}. The sample labelling complexity is the effort required to provide the ground truth labels for each sample. For NID datasets, the sample complexity is in the range of millions. Thanks to automation today, the security research community has paid less attention to the most relevant problem of verifying the correctness~\cite{liu2022error} of the auto-labeling process. The labelling budget also affects data annotation. It is difficult for security practitioners to fix the labelling budget due to the variations in the collected data and subsequent labelling efforts, which depend on the specific environment. Eventually, the gap between academic research and real-world security practices is the reason for a lack of established systematic guidelines for the data annotation process in the cybersecurity domain. One way to reduce the reliance on labeled data is to employ the semi-supervised continual learning (SSCL) methods due to their ability to work with limited annotated data. Similar to prior works~\cite{amalapuram2023augmented}, our work also considers the NID problem as a binary classification problem in a domain incremental learning setting.

\par \textbf{Open-world learning in CL-based NIDS:} One of the long-standing goals of the security domain is to identify the unseen attacks (\textit{a.k.a.} zero-day attacks) with high \textit{confidence} by leveraging the learned knowledge. Specifically, the goal is to distinguish zero-day attacks from normality shift detection. However, existing CL-based NIDS detection performance degrades while identifying unseen attacks. This degradation occurs whenever the distribution of the unseen data during inference differs from the training data distribution. While continual learning strategies strive to reduce the effect of CF on the seen tasks, their performance effectiveness remains questionable on unseen tasks.

\begin{table}[H]
\centering
  \caption{The performance decay of the classifier on the attack class detection on the CICIDS 2017 and CSE-CICIDS-2018 datasets. The reported values are area under time (AUT) computed on the precision-recall area under the curve values of the attack class (AUT (A)).} 
  \label{tab:unseen_aut_demo}
  \centering
  
  \begin{tabular}{llll}
  
    \midrule
    & \multicolumn{2}{c}{AUT (A)}    \\
    \cmidrule(lr){2-3} 
   Dataset  & Seen traffic & Unseen traffic \\
   \midrule
   CICIDS-2017~\cite{liu2022error} & 0.985 & 0.506\\
   CSE-CICIDS-2018~\cite{liu2022error} & 0.999 & 0.157 \\  
   \bottomrule
   
  \end{tabular}
\end{table}
We demonstrate this empirically for network intrusion detection tasks using two representative datasets, CICIDS-2017~\cite{liu2022error} and CSE-CICIDS-2018~\cite{liu2022error}. The CICIDS-2017 and CSE-CICIDS-2018 datasets consist of traffic spread over five and ten days, respectively. 
Initially, a classifier (multilayer perception) is trained on the first two days of traffic in a fully supervised fashion. This model is then tested on the remaining days of traffic. The performance evaluation is measured using the area under the time (AUT) computed on the precision-recall area under the curve (PR-AUC) for attack class data. From the results shown in Table~\ref{tab:unseen_aut_demo}, we observe that the detection performance of the attack class data on the unseen traffic deteriorates over time. 

\par Motivated by the label scarcity and performance decay of the classifier on the unseen traffic, we formulate the open-world learning problem as a semi-supervised continual learning method that requires limited annotated data. 
The key contributions of this work can be summarized as follows: 
\begin{enumerate}
    \item The formulation of an open-world learning problem for security applications as a semi-supervised CL problem, dubbed SOUL, incorporates test-time distributions with scarce label data for unseen network traffic.
    \item Operating under the assumption of no prior knowledge about these unseen tasks, the method produces high-confidence labels in conjunction with buffer memory and model confidence when the unseen traffic bear similarities to the seen traffic. (Section~\ref{sec:train_unseen}).
    \item An evaluation of the proposed method on various benchmarks (Section~\ref{sec:results}) which demonstrates that its performance is on par with the fully supervised continual learning baselines using a maximum of 20\% labeled data and reduces the data annotation effort in the range of 11 to 45\% on the unseen data.
\end{enumerate}

The rest of the paper is organized as follows. Section~\ref{sec:related_work} describes the related work. Section~\ref{sec:supplement} describes the background details and introduces the notion of orthogonal projections and how to compute them. Section~\ref{sec:sscl_for_nid} describes the SSCL notations for the NID problem. Section~\ref{sec:methodology} explains the proposed methodology, training scheme, and pseudo-code. Section~\ref{sec:results} details the experimental setting, performance results, and ablation studies. 

\section{Related Work}
\label{sec:related_work}
 Network intrusion detection is a form of malicious activity detection in communication network traffic. Malicious activities can include unauthorized access to sensitive information, making target systems unavailable to flooding requests (denial-of-service attacks), etc. We now review the recent related work on NIDS in the context of CL and other incremental learning setting.

\par \textbf{Continual learning for NID:} The first experiment involving CL algorithms on NIDS datasets appeared in~\cite{8682702}. Although the authors proposed an extension method based on a variational autoencoder and generative replay for anomaly detection in the context of computer vision tasks, they evaluated the proposed method on the MNIST and NSL-KDD datasets. Later, the authors of ~\cite{amalapuram2022continual} studied the suitability of different families of CL approaches to the NID problem and showed that memory replay-based methods are more feasible. ~\cite{10020865} proposed a new buffer memory management technique based on the progressive filling algorithm for CL-based NID to handle data imbalance problems that store more attack samples in memory than random sampling. The efficacy of this method was evaluated using the NSL-KDD dataset. Recently,~\cite{amalapuram2023augmented} proposed a new MR method dubbed ECBRS to handle the severe class imbalance in NID datasets and PAPA methods to reduce the total training time. These methods were evaluated using eight publicly available NID datasets.  However, all of these methods are designed for supervised learning, requiring large amounts of annotated data, with the experiments conducted under a closed-world assumption. In contrast, the proposed method operates in a semi-supervised learning setting emphasizing open-world learning.

\textbf{Other incremental learning methods:} INSOMNIA~\cite{andresini2021insomnia} is the first work of its kind to address concept drift (CD) in network intrusion detection using a semi-supervised learning setting. To reduce manual labeling overhead, it employs a nearest centroid neighborhood classifier to estimate labels. However, under extreme drift in network traffic, this method may be prone to self-poisoning, affecting the quality of the self-labeling process. OWAD~\cite{hananomaly} addresses CD as a normality shift adaptation using an anomaly detection technique. The idea is to periodically adapt to CD in benign traffic, thereby indirectly managing shifts in attack traffic. This method may fail to distinguish whenever the evolved attack traffic is similar to benign. Trident~\cite{10.1145/3589334.3645407} incrementally adds multiple one-class classifiers to identify and adapt to new, unseen classes. However, with the ever-increasing number of attacks, adding new one-class classifiers may become a bottleneck and ultimately increase the overall complexity of the system. To reduce the manual labeling effort required to adopt to CD, ReCDA~\cite{yang2024recda} combines self-supervised representation enhancement and weakly-supervised classifier tuning. The proposed method uses perturbation techniques as a part of self supervision, however such perturbation in feature space (as opposed to problem space) of network traffic recently received a criticism~\cite{apruzzese2024when}. In contrast to existing methods, our proposed approach does not rely on perturbations, does not increase model complexity, and adapts to concept drift in both benign (normality shift) and attack traffic.
  \section{Background}
    \label{sec:supplement}
\par This section delves into the concept of orthogonal projections in CL. It then provides a detailed explanation of how to compute these projections using gradient projection memory (GPM).
\label{subsec:gpm_update}
\par For orthogonal projections, prior works~\cite{farajtabar2020orthogonal,chaudhry2018efficient,lopez2017gradient} store important gradient directions or exemplars to obtain reference gradient directions, however, these methods require higher storage requirements. In contrast, GPM leverages the fact that stochastic gradient descent (SGD) updates lie in the span of the input data~\cite{zhang2017understanding}. Intuitively, it is represented using Equation~\ref{eq:sgd_span}.
\begin{equation}
    \nabla_{\theta}\mathcal{L} = (\theta^{T} \mathbf{x} - \mathbf{y}) \mathbf{x}^{T} = \delta \mathbf{x}^{T},
    \label{eq:sgd_span}
\end{equation}
where $\nabla_{\theta}\mathcal{L}$ is the gradient of the objective loss function $\mathcal{L}$ with model parameters $\theta$, $\delta$ is the error between the ground truth and predicted value and $\mathbf{x}$ is an input data point.
\par Based on the prior intuition, the gradient space is split into two orthogonal sub-spaces: core gradient space (CGS) and residual gradient space (RGS). While learning a new task, taking gradient steps toward CGS will yield higher interference with past tasks, whereas orthogonal steps to CGS will yield minimal interference. Thus, after training with each task, CGS is stored in memory. To avoid the high storage requirements, representations (activations) of all tasks are stored instead of CGS, as the input data spans the gradient space. However, storing past task activations requires higher storage. This problem is mitigated by storing the basis of all the representations by using a low-rank matrix constructed using the singular value decomposition (SVD) algorithm. The low-rank matrix is iteratively updated with a new basis vector after training with a new task; such a matrix is known as \textit{gradient projection memory}.

\par \textbf{How to compute representations for the first task?}: The representations denote the activation function values at each layer, aligned with the input. For instance, an activation function is sigmoid, tanh, rectified linear unit (ReLU), and leaky ReLU. Let's denote the learning network as $f_{\theta}$ with $k$ layers and the activation function by $\sigma(.)$. The resulting output, $\mathbf{x}^k$ (corresponds to input $\mathbf{x}$) represents the activations at the $k$-{th} layer. The corresponding representations at each layer can be recursively written using Equation~\ref{eq:activations}.

\begin{equation}
  \begin{gathered}   
  \mathbf{x}^{1} = \sigma(f_{\theta_1}(\mathbf{x}))\\
  \mathbf{x}^{2} = \sigma(f_{\theta_2}(\mathbf{x}^1))\\
  \vdots\\
   \mathbf{x}^{k} = \sigma(f_{\theta_k}(\mathbf{x}^{k-1}))
    \end{gathered}    
    \label{eq:activations}
  \end{equation}
  whereas $f_{\theta_k}$ are the parameters of the $k^{th}$ layer, $\mathbf{x}^{k-1}$. Initially, after training with data from the first task, a randomly selected exemplar is selected per task class. These exemplars are used for the learner's forward pass step; during this, activations are accumulated at the respective layer. All these activations are collected to maintain in a matrix known as the \textit{representation matrix} corresponding to the task. The representation matrix of the network's $k^{th}$ layer for the first task is constructed by column-wise concatenating the activations corresponding to $n_{c}$ randomly selected exemplars. The matrix is denoted as $R_1^{k}$ 
  \begin{equation}
      R^k_{1} = [ \mathbf{x}^{k}_1,\mathbf{x}^{k}_2,\cdots,\mathbf{x}^{k}_{n_c}]
  \end{equation}
  where $\mathbf{x}^i_j$ is the activation (representation) of the $j^{th}$ example at the $i^{th}$ layer.
  \par \textbf{Finding the basis vectors of representation matrix:}  The representation matrix size grows with the size of the network and affects the total training time. To overcome this issue, basis vectors of the representation are computed using the SVD algorithm, in which the representation matrix as  ${R}^{k}_{1} = {U}^{k}_1 \Sigma^{k}_1 ({V}^{k}_1)^{T}$ to obtain the $q-$rank approximation $({R}^{k}_{1})_q$ satisfying the below-mentioned criteria:
  \begin{equation}
  \begin{gathered}      
   ||({R}_{1}^{k})_{q}||^2_{F} \geq \delta^{k} ||{R}^{k}_{1}||^2_{F}
    \end{gathered}
    \label{eq:k-rank-approx}
  \end{equation}
The first $q-$left singular vectors of the matrix $U$ will be treated as the basis vectors of the $k^{th}$ layer and stored in the projection memory. Similarly, this process is repeated for all the remaining layers for the first task, and the corresponding basis vectors are added to the memory. The GPM for the first task is given as  
\begin{equation}
  \begin{gathered}   
  \mathcal{M}_{gpm}^1=\{(\mathcal{M}^k)_{k=1}^{K}\}\\
  \mathcal{M}^{k}= [\mathbf{u}_{1,1}^{k},\mathbf{u}_{2,1}^{k},\cdots,\mathbf{u}_{q,1}^{k}]
    \end{gathered}
    \label{eq:gpm_first task}
  \end{equation}
  where $\mathbf{u}_{j,1}^{k}$ is the $j^{th}$ left singular vector of the $k^{th}$ layer for the first task.

\par \textbf{Finding projection memory for subsequent tasks (from 2 to t):} From the second task onwards, the computed gradient is projected orthogonally to the previous task projection memory before the backpropagation operation. Let us assume that the gradient computed concerning the loss function of the second task for the $k^{th}$ layer is  $\nabla_{\theta^{k}}\mathcal{L}_{2}$. The orthogonal projection of the gradient concerning the first task uses Equation~\ref{eq:gpm-orthogonal}. Hence, the residual gradient component lies in the space orthogonal to the core gradient space.
  \begin{equation}
  \begin{gathered}      
   \nabla_{\theta^{k}}\mathcal{L}_{2} = \nabla_{\theta^{k}}\mathcal{L}_{2} - (\nabla_{\theta^{k}}\mathcal{L}_{2}) \mathcal{M}^{1}_{gpm} (\mathcal{M}^{1}_{gpm})^{T}
    \end{gathered}
    \label{eq:gpm-orthogonal}
  \end{equation}

\par where $\mathcal{M}^{1}_{gpm}$ is the gradient projection memory for the first task. Hence, the projected gradient lies in the residual space. 
\par \textbf{Eliminating redundant basis vector:} Before applying SVD to compute the basis vector, it is important to eliminate the redundant vectors between the representations for the second task and the basis vector present in the GPM. Mathematically, redundant vector removal operation is computed using the following equation.

\begin{equation}
  \begin{gathered}      
   \hat{{R}^{k}_{2}} ={R}^{k}_{2} - \mathcal{M}^{1}_{gpm} (\mathcal{M}^{1}_{gpm})^{T}{R}^{k}_{2}
    \end{gathered}
    \label{eq:common-bases}
  \end{equation}
  \par where, ${R}^{k}_{2}$ is the representations of the $k^{th}$ layer for second task. Now, SVD is applied to find the $k-$rank approximations basis vectors that are added to $\mathcal{M}_{gpm}$. This process is continued for all layers for the remaining tasks. Thus, the GPM will span the gradient space for past tasks.
\begin{equation}
  \begin{gathered}      
  \text{span} \{\nabla_{\theta}\mathcal{L}_1, \nabla_{\theta}\mathcal{L}_2, ..., \nabla_{\theta}\mathcal{L}_{t-1}\} = \mathcal{M}_{gpm}
    \end{gathered}
    \label{eq:gpm_whole_tasks}
  \end{equation}

\section{Semi-Supervised Continual Learning for NID}
\label{sec:sscl_for_nid}
\par In this section, we formally define the problem of semi-supervised continual learning (SSCL) in the open-world learning (OWL) setting. The training data arrives as a sequence of $\mathcal{T}$ tasks, and the data associated with the task `$t$' is represented as $\mathcal{D}^t$, where $t \in \{1,2,\dots, \mathcal{T}\}$. For training in OWL, the entire dataset is partitioned into two sets as seen and unseen tasks. Since we operate in a semi-supervised setting, only partially labeled data is available for each seen task `$t$'. Thus, $\mathcal{D}^t = \{\mathcal{D}^{t}_{l}, \mathcal{D}^{t}_{u}\}$, where $\mathcal{D}^{t}_{l}$ and $\mathcal{D}^{t}_{u}$ are the labeled and unlabeled exemplars of the task $t$, respectively, and $|\mathcal{D}^{t}_{l}| \ll |\mathcal{D}^{t}_{u}|$.  On the contrary, for the task $t^{\prime}$ in unseen tasks, only unlabeled data is available {i.e.,} $\mathcal{D}^{t^{\prime}} = \{ \mathcal{D}^{t^{\prime}}_{u}\}$. Similar to the previous work~\cite{amalapuram2023augmented}, this work also operates in the domain incremental learning setting, where the label space ($0$ or $1)$ remains the same across all the tasks.

\par Assuming a general continual learning setting, let $f_{\theta}$ be a solver function parameterized by $\theta$. The solver function is trained using a sequence of tasks from the dataset. At any point in time during training, $f_{\theta}$ will access only the dataset of particular task $t$, which is $\mathcal{D}^t$. However, limited access is provided to the past task's samples to store in buffer memory, which helps mitigate catastrophic forgetting. During the training on $\mathcal{D}^{t} (t>1)$, the pseudo labels for $\mathcal{D}^{t}_{u}$ of the seen tasks are generated using the solver function trained on the previous batch of samples. The training process of the task $t^{\prime}$ of the unseen tasks is discussed in Section~\ref{sec:train_unseen}.


\section{Methodology}
\label{sec:methodology}
The success of continual learning for network intrusion detection in the open-world setting depends on the classifier's precision in identifying novel cyber attacks by leveraging the learned knowledge. When the classifier is trained under full supervision, it is required to annotate all the unseen samples on which the classifier is uncertain. However, data annotation is a complex process in the security domain, requiring superior domain knowledge to label thousands of samples. To alleviate these data annotation issues, we operate in the semi-supervised continual learning setting that requires limited labeled data to adapt to newer tasks.
\par We first describe the threat model and assumptions about the environment and the classifier the proposed method requires. Then, we present a motivating example for using the orthogonal projections in the SSCL setting. Eventually, the training process of the seen and unseen tasks using the proposed SOUL method is discussed.
\begin{figure*}
  \begin{center}
    \includegraphics[scale=0.25]{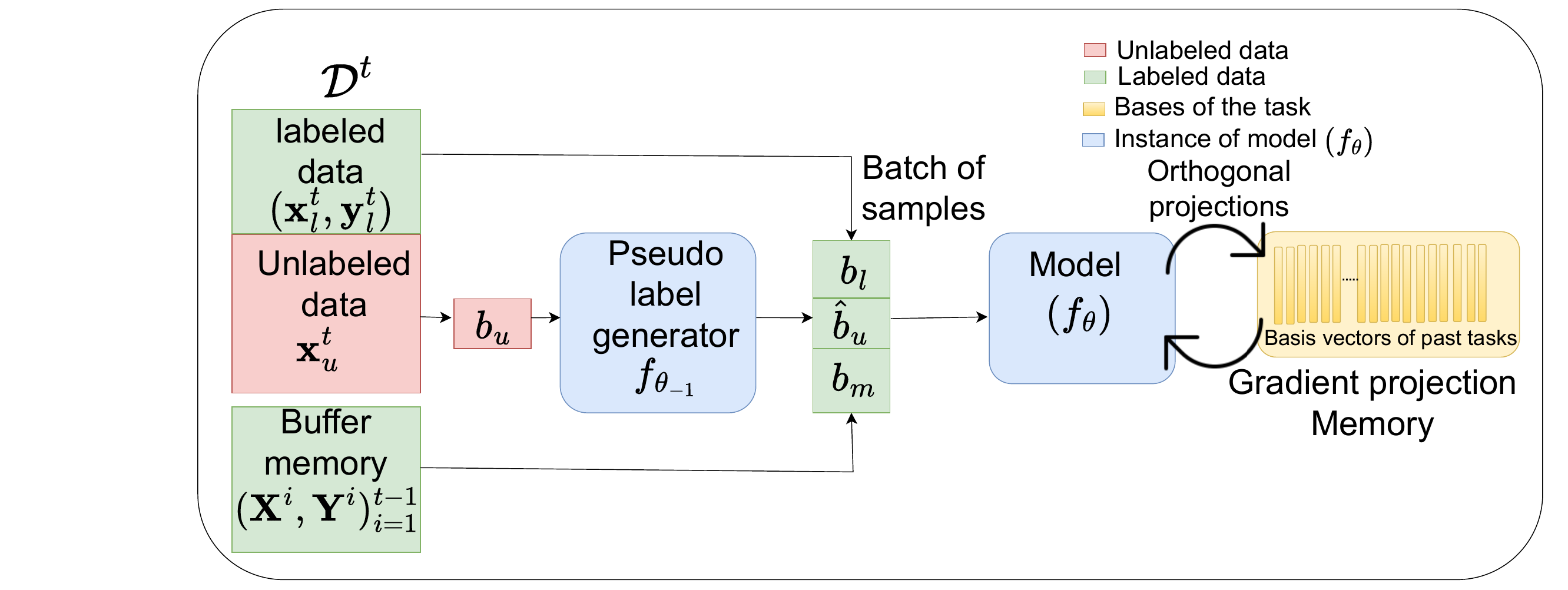}
  \end{center}
  \caption{Graphical illustration of the end-to-end training process of the proposed SOUL method on each of the seen tasks.}
  \label{fig:train_seen_task}
  \end{figure*}
\subsection{Threat model and assumptions}
In this work, we make the following assumptions about the attacker's knowledge, the environment, the ML-based defense mechanism, and the labelling assumption.
\par \textit{Attacker knowledge}: The attacker aims to steal sensitive information and gain unauthorized access to the host machines in an enterprise network. The attacker does not know the datasets or algorithms used in the ML-based defense mechanism deployed by the enterprise. 
\par \textit{Environment:} We assume \textbf{a network environment} that spans hundreds of host machines. Traffic generated by hosts is captured using packet captures (PCAPS) and translated into NetFlow records using aggregating software. These records are analyzed to detect different attacks by NIDS. 
\par \textit{ML-based classifier:} The ML classifiers can quickly become obsolete due to the concept drift in the benign and malicious traffic. During retraining, access is restricted to a subset of past training data.
\par \textit{Labeling assumptions and labelling budget}: The ML classifier can generate labels for new/unseen network traffic (flow records) similar to previously seen traffic. However, security analysts will also create labels for partial data representing new threats where the model is highly uncertain. The proposed method doesn't consider the budget for analysts' labelling efforts.


\subsection{Leveraging orthogonal projections as complementary inductive biases: An empirical motivation}
\label{subsec:work-motivation}
\par The success of machine learning in domains such as computer vision and natural language processing can be attributed to inductive bias (IB), which is a set of assumptions made by a learning algorithm to improve the generalization performance. For example, the IB for images is spatial correlation. Thus, the widely used Convolutional Neural Network (guided by these IB) outperforms human performance in various visual cognitive tasks.

ML-based security researchers focus on learning classifiers from hand-crafted features, and the possible IB is handling the \textit{class imbalance} (CI) when trained in a supervised/semi-supervised fashion. Many researchers formulated this IB in different forms, such as resampling (SMOTE)~\cite{fernandez2018smote}, cost-sensitive learning~\cite{thai2010cost}, and using buffer memory in the CL framework~\cite{chrysakis2020online}. Our empirical findings suggest that \textit{handling CI using buffer memory may not be effective IB} in CL settings to improve the performance of the attack class detection. 

\begin{table}[!htb]
\centering
    \caption{The performance results of the SSCL method on standard NID benchmark datasets. The method SSCL+GPM indicates SSCL methods uses orthogonal projections via gradient projection memory. The attack and benign class  PR-AUC values are denoted as PR-AUC (A) and PR-AUC (B). The improved performance results of the minority class are marked in bold.}

    \begin{tabular}{llll}
    
    \cmidrule(lr){1-4}
         Dataset& Method & PR-AUC (A) & PR-AUC (B) \\
     \midrule  
         CICIDS-2017 & SSCL & 0.651 & 0.959\\   
          & SSCL+GPM & \textbf{0.935} & 0.995\\
         \midrule
         CSE-CICIDS-2018 & SSC & 0.424 & 0.970\\   
          & SSCL+GPM & \textbf{0.999} & 0.999\\
        
         \bottomrule
    \end{tabular}
    \label{tab:discriminatio_demo}
    \end{table}
\par \textit{Motivating example:} We experiment on NID datasets to understand the efficacy of the orthogonal projections (using the gradient projection memory (GPM)~\cite{saha2020gradient} method) in an SSCL setting. The SSCL method used here inherently handles CI using memory. The SSCL method trains the classifier task-wise; each task contains labeled and unlabeled data. The pseudo labels for unlabeled data are generated using the classifier trained on the previous batch of samples, and a buffer memory of past samples is used to mitigate CF. These experiments follow a closed-world setting, where test and train data are from the same distribution. The performance results are presented in Table~\ref{tab:discriminatio_demo}. We observe that the minority (attack) class performance is significantly improved across all the benchmarks (marked in bold) and by a maximum of 2x (in the case of CSE-CICIDS-2018). Further, it is interesting to note that the PR-AUC values of the majority class also improved across all the datasets. Thus, augmenting with orthogonal projections-based methods will improve attack class detection performance. The detailed steps for obtaining orthogonal projections using gradient projection memory are presented in Section~\ref{subsec:gpm_update}.
\begin{figure*}
  \begin{center}
    \includegraphics[scale=0.15]{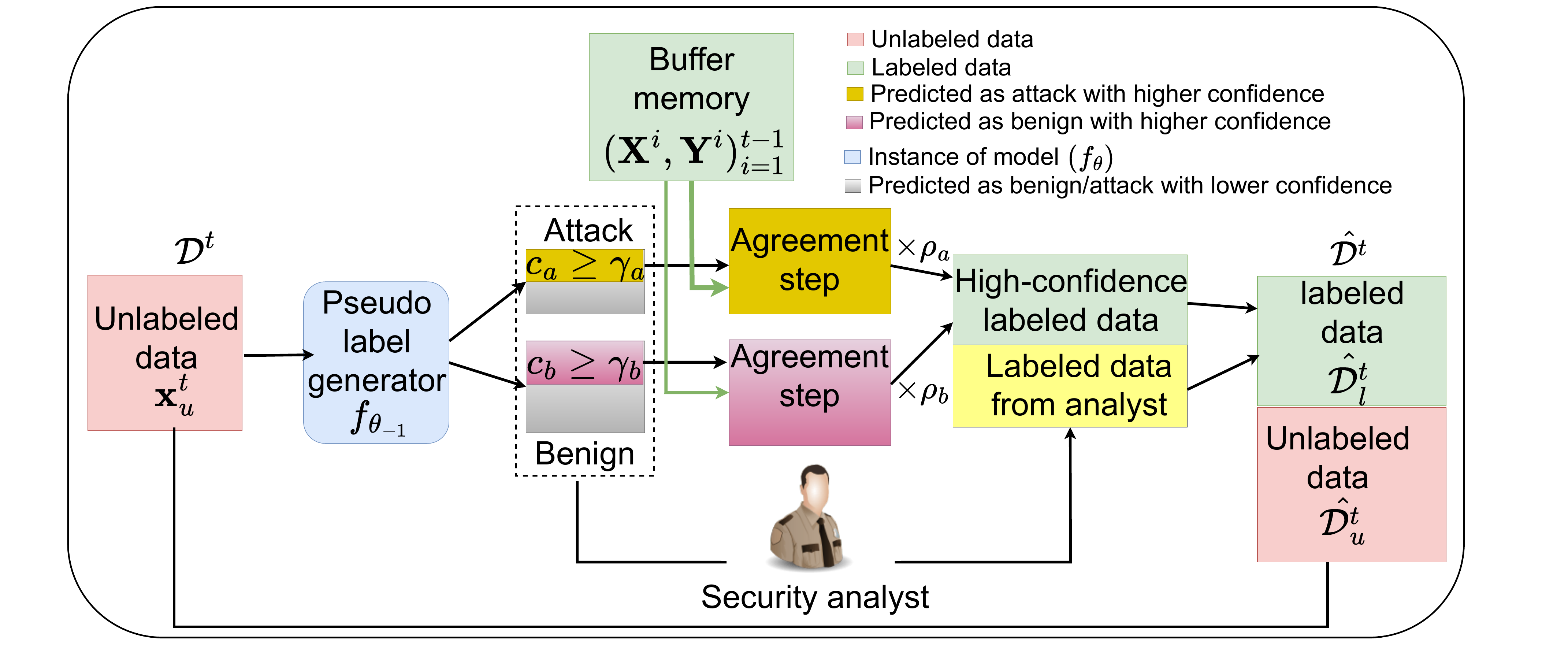}

  \end{center}
  \caption{Graphical illustration of the end-to-end training process of the proposed SOUL method on each unseen task in open-world learning.}
  \label{fig:train_unseen_task}
  \end{figure*}
\par In a nutshell, CI hinders the detection of minority (attack) class instances within the SSCL framework. Addressing CI as an inductive bias during classifier training is insufficient to enhance attack class detection. Our empirical findings indicate that combining SSCL with a CI handling mechanism and orthogonal projections improves the detection performance of the attack class. 

\subsection{Training scheme}
\par The classifier/model is first trained on seen tasks in an SSCL setting. Following this, the classifier is trained on unseen tasks in which high-confidence labels (limited) for unlabeled data are predicted in conjunction with the learned model and buffer memory. These partially labeled data are used to learn the unseen task in an SSCL setting.  Thus, OWL is formulated as an SSCL problem.
 \subsubsection{Training with seen tasks} 
 \label{sec:train_known_task}
 The training process on seen tasks contains partially labeled data per task and a large pool of unlabeled data. The training process follows domain incremental learning~\cite{amalapuram2023augmented} to solve a binary classification problem. During the training of each task ($t$), the data for each batch of samples arrives from three sources. They are labeled data (of size $b_l$), unlabeled data (of size $b_u$) from the task $`t$', and labeled data (of size $b_m$) from the buffer memory. 

The labels for the unlabeled data are generated using a knowledge distillation technique, specifically a teacher-student~\cite{gou2021knowledge} model. The teacher model is a classifier trained until the previous batch of samples. 

The usage of the teacher-student model is based on the intuition that every batch of samples contains labeled data of the current task, past tasks (sampled from buffer memory), and further orthogonal gradient projections helps to maintain current and past knowledge are useful to generate labels for the unlabeled data. 
The overall objective function is as follows.

\begin{align}
\label{eq:whole_loss_2}
\mathcal{L}(\mathcal{D}^t) &= \sum_{\substack{(\mathbf{x}_{l}^{t},\mathbf{y}_{l}^{t}), \mathbf{x}_{u}^{t}\sim \mathcal{D}^{t}\\  (\mathbf{x}_{m},\mathbf{y}_{m}) \sim \mathcal{M}}} 
\ell_{c}(f_{\theta}(\mathbf{x}_{l}^{t} \cup \mathbf{x}_{m} \cup \mathbf{x}_{u}^{t}), \mathbf{y}_{l}^{t} \cup y_{m} \cup \hat{y}_{u}) \nonumber \\
& \quad  \hspace{25mm} +   \ell_{dist}(f_{\theta}(\mathbf{x}_{u}^{t}), f_{\theta_{-1}}(\mathbf{x}_{u}^{t})),
\end{align}

\par where $(\mathbf{x}_{l}^{t},\mathbf{y}_{l}^{t})$ and $\mathbf{x}_{u}^{t}$ are the labeled and unlabeled data drawn from the current task, $(\mathbf{x}_{m},\mathbf{y}_{m})$ are sampled from memory, $\ell_c(.)$ is the classification loss, $\ell_{dist}(.)$ is the distillation loss, and $f_{\theta_{-1}}$ is the teacher model. After computing the gradients of the objective function with respect to the model parameters, the gradients are orthogonally projected into GPM before applying the back propagation algorithm. The graphical illustrations of the training process is presented in Figure~\ref{fig:train_seen_task}. GPM is updated using the procedure described in Sections~\ref{subsec:gpm_update} after finishing the training with the current task. 

  \subsubsection{Training with unseen tasks in OWL} 
  \label{sec:train_unseen}The training process on unseen tasks involves obtaining the labels for a subset of unlabeled data, 
  later trained using the SSCL setting described in Section~\ref{sec:train_known_task}. Initially, using the classifier ($f_{\theta_{-1}}$) trained on seen tasks, the prediction confidence and pseudo labels (using the argmax over probability distribution) of the unlabeled data are computed. The entire unlabeled data is partitioned into two classes based on the generated pseudo labels, and each partition is sorted based on the prediction confidence values in non-increasing order. During seen task training, the class imbalance ratio (CIR) for each task is recorded, and the mean CIR value is used as the expected CIR for unseen tasks. 
  We empirically validate the effectiveness of using mean CIR as a substitute for unknown CIR of unseen tasks in Section~\ref{mean_cir_effectiness} and find that it works well on unseen tasks. Using this CIR per each class partition, the top $d_l$ amount of labeled data is chosen with prediction confidence ($c_{a},c_{b})$ above the pre-determined threshold ($\gamma_{a},\gamma_{b}$ for pseudo attack and benign classes). This step will ensure a similar labeled data ratio is preserved in unseen tasks. These chosen samples are the top samples that belong to the respective class partitions based on the models' prediction confidence. The threshold value is set to 0.9 (or higher) to consider the high confidence assigned labels by the classifier.
  \par Next, in the agreement step, we compare the chosen samples with the known exemplars from the previous tasks to find the \textit{closeness} with the old tasks. The intuition behind the agreement step can be viewed as \textit{revalidation} of the generated pseudo labels to minimize the false positives (FPs). In other words, we compute the distance between the past task samples and chosen samples (of unseen tasks that are pseudo labeled by classifier) to obtain the labels using the majority vote among the closest classes of past tasks. The predicted pseudo-labels that agree with the labels generated using closeness metrics are considered to ensure self-labeling quality. This step ensures the model's prediction consistency in identifying known exemplars using the closeness metric. 
  Our work uses a buffer memory for closeness computation. The memory contains a subset of exemplars from past tasks that act as a representative of all previous tasks. 
 
\par \textbf{Handling FPs using majority voting scheme}: For each chosen sample (of unseen tasks), we calculate cosine distance against all memory samples. The samples from memory whose cosine distance is less than \textit{$c_d$} are selected. Among these selected samples, the label corresponding to the majority percentage that exceeds a specified threshold is considered the final label.  This approach effectively handles the variance introduced by low-similarity memory samples, enhancing the performance of the SOUL method, as demonstrated through ablation studies in Section~\ref{sec:ablation-study}.

  We randomly select $\rho_a$ and $\rho_b$ percentages from the agreed-upon set of samples for attack and benign classes. These selected samples are then included as labeled exemplars. Here, $\rho_a$ and $\rho_b$ are known as the agreement-fraction values that are computed from the \textit{seen} tasks during their training process (Section~\ref{sec:train_known_task}). We perform a similar agreement step on the seen tasks using the labeled data to compute what fraction of model predictions on the seen task agree with their ground truth value. A weighted average of the current fraction and past task agreement fraction gives $\rho_a$ and $\rho_b$. 
 
  \par Ultimately, if the number of self (model)-labeled samples selected through model predictions is insufficient, a security analyst labels the necessary samples. This intuitively means that if the number of samples selected does not meet the desired labeled data ratio, analysts will provide labels for the remaining samples. The proposed approach doesn't substitute for the role of security analysts but rather complements it, assisting them in generating high-quality labels based on model consistency. In extreme situations, especially in the presence of higher distribution shifts, no ML-based method can generate high-quality labels, and the proposed method is no exception. In such cases, analysts alone can provide ground truth labels. After acquiring the required labeled data, the current unseen task is trained using the training process of a known task in the SSCL setting (described in Section~\ref{sec:train_known_task}). The graphical illustration of the training process is presented in Figure~\ref{fig:train_unseen_task}. The pseudocode of the training process of the seen and unseen tasks is available in Section~\ref{append:sec_pesudocode}.
\par \textbf{Effectiveness of SOUL on unseen attacks:}
When novel attacks that are dissimilar to previous ones occur, the SOUL method employs a disagreement step to identify these novel samples as \textit{unknown}, as follows: disagreement occurs whenever there is a mismatch between the classifier's pseudo-label and the majority vote labeling for the novel sample. This disagreement step can also be viewed as a way of estimating the classifier's uncertainty in labeling a novel sample; during this process, the classifier's pseudo-label is discarded. Intuitively, this means that a novel sample is identified as an unknown category (neither benign nor an attack). In this situation, the SOUL method relies on the label generated by the analyst for the novel sample. This approach aligns with established practices in the security and ML community for identifying uncertain novel samples as unknown using classifier confidence~\cite{sun2024exploring,zhu2024survey,291253} and involving human expertise (or an oracle) for labeling an unknown novel sample~\cite{291253,joseph2021towards}.
\subsection{Pseudo-code of the proposed SOUL method}
\label{append:sec_pesudocode}
\par This section describes the pseudo-code for the proposed SOUL method. The training process starts with training the first task containing only labeled data using Algorithm~\ref{alg:training} (steps 4 to 9) and buffer memory reorganization at step 16.
\begin{algorithm}[htb!]
\caption{Training algorithm of the proposed  SOUL method}
   \label{alg:training}
\begin{algorithmic}[1]
   \STATE {\bfseries Inputs:} 
   the sequence of $\mathcal{T}$ tasks $\{1,2,\cdots,\mathcal{T}-1,\mathcal{T}\}$, number of seen/known tasks trained in closed-world setting $c$, dataset of seen task `$t$' ($t <=c $) is $\mathcal{D}^{t} = \{\mathcal{D}^{t}_{l},\mathcal{D}^{t}_{u}\}$ (labeled and unlabeled portions), dataset of unseen task `$t$' ($t > c$) is  $\mathcal{D}^{t} = \{\mathcal{D}^{t}_{u}\}$ (only unlabeled data),
   batch size $b$, model $f_{\theta}$, buffer memory $\mathcal{M}$, and memory partition factor $\mathcal{M}_{alloc}$

    \STATE {\bfseries Output:} $f^{\mathcal{T}}_{\theta}$ model trained on all $\mathcal{T}$ task
    \FOR{each task `$t$'}
        \IF{$t = 1$ (first task)}
        \WHILE {$\mathcal{D}^{t}_{l}$ is non-empty}
        \STATE sample labeled data $\mathcal{B} \sim \mathcal{D}^{t}_{l}$, where  $|\mathcal{B}| = b $ 
        \STATE compute gradient $\nabla f^{t}_\mathcal{B}$ of classification loss $\mathcal{L}_{t}$ on  $\mathcal{B}$
        \STATE update $f_{\theta}^{t}$ using the  $\nabla f^{t}_\mathcal{B}$
        \ENDWHILE

        \ELSIF {$t \leq c$ (closed-world tasks)}
            \STATE $f^{t}_{\theta}$ = Semi-supervised continual learning ($\mathcal{D}^{t}_{l}, \mathcal{D}^{t}_{u}$) using the Algorithm~\ref{alg:semi-supervised}
            
        \ELSE 
            \STATE $\mathcal{\hat{D}}^{t}_{l}, \mathcal{\hat{D}}^{t}_{u}$ = Open world labeling ($\mathcal{D}^{t}_{u}$) using the Algorithm~\ref{alg:owl-labeling}
            \STATE $f^{t}_{\theta}$ = Semi-supervised continual learning ($\mathcal{\hat{D}}^{t}_{l}, \mathcal{\hat{D}}^{t}_{u}$) using the Algorithm~\ref{alg:semi-supervised}
        \ENDIF
        \STATE reorganize the buffer $\mathcal{M}$ with the $\mathcal{M}_{alloc}$ using the labeled data $\mathcal{D}^t_l$ (or $\mathcal{\hat{D}}^t_l$ for unknown tasks)
    \ENDFOR
    \RETURN $f^{\mathcal{T}}_{\theta}$  
\end{algorithmic}
\end{algorithm}
\begin{algorithm}[htb!]
    \caption{Semi-supervised continual learning}
    \label{alg:semi-supervised}
    \begin{algorithmic}[1]
        \STATE {\bfseries Input:} 
         dataset for task `$t$' (with labeled and unlabeled portions) $\mathcal{D}^{t} = \{\mathcal{D}_{l}^{t}, \mathcal{D}_{u}^{t}\}$,
        labeled dataset from previous task $\mathcal{D}^{t-1}_{l}$,
        model trained till the previous task `$t-1$' $f^{t-1}_{\theta}$,
        batch size $b$, 
        buffer memory $\mathcal{M}$,
        no. of samples drawn from $\mathcal{M}$: $b_m$ ($ \leq b$),
        gradient projection memory $\mathcal{M}_{gpm}$,
        labeled data ratio $r$,

        \STATE {\bfseries Output:} 
        $f^{t}_{\theta}$ - model trained until task `$t$'
        
        \STATE sample $n_c$ exemplars from the attack (minority) class of $\mathcal{D}^{t-1}_{l}$ 
        \STATE compute activations ${A}^{t-1}$ of $n_c$ samples using $f^{t-1}_{\theta}$ using Eq~\ref{eq:activations}
        \STATE compute ${A}_k^{t-1}$ using SVD (using Eq~\ref{eq:k-rank-approx}), add it to $\mathcal{M}_{gpm}$ (using Eq~\ref{eq:gpm_whole_tasks})

        \WHILE{$\mathcal{D}^{t}$ is non-empty}
             \STATE sample labeled data from buffer memory $\mathcal{B}^{m}_{l} \sim \mathcal{M}$, where $|\mathcal{B}^{m}_{l}|= b_m$
             \STATE  $b_{rem} =b - b_{m} $
             \STATE sample labeled data from current task $\mathcal{B}_{l} \sim \mathcal{D}^{t}_{l}$, where $ |\mathcal{B}_{l}| = b_l = b_{rem} \times r$ 
            \STATE  $b_{u} = b_{rem}-b_l $
            \STATE sample unlabeled data from current task $\mathcal{B}_{u}   \sim \mathcal{D}^{t}_{u}$, where $|\mathcal{B}_{u}|$ = $b_{u} $ 
            \STATE generate pseudo labels $\hat{\mathcal{B}}_{u}$ (for the samples $\mathcal{B}_{u}$) using $f_\theta^{t-1}$
            \STATE $\mathcal{B} = \hat{\mathcal{B}}^{m}_{l} \cup \mathcal{B}_{l} \cup \hat{\mathcal{B}}_{u} $
            \STATE compute gradient $\nabla f^{t}_\mathcal{B}$ of loss $\mathcal{L}_{t}$ (computed using Eq~\ref{eq:whole_loss_2}) on  $\mathcal{B}$    
            \STATE compute $\nabla^{\prime} f^{t}_\mathcal{B}$, orthogonal projection of  $\nabla f^{t}_\mathcal{B}$ on $\mathcal{M}_{gpm}$ using Eq~\ref{eq:gpm-orthogonal}
            \STATE update $f_{\theta}^{t}$ using $\nabla^{\prime} f^{t}_\mathcal{B}$ 
        \ENDWHILE
        \RETURN $f_{\theta}^{t}$
    
    \end{algorithmic}
\end{algorithm}
\par From the second seen task to the last seen task, Algorithm~\ref{alg:training} at step 11 invokes the semi-supervised continual learning algorithm (Algorithm~\ref{alg:semi-supervised}),  which computes the gradient projection memory on the past task (steps 3 to 5 of Algorithm~\ref{alg:semi-supervised}) and constructs a batch of samples from labeled, unlabeled data of the current task,  and labeled data from memory $\mathcal{M}$ to compute the gradients concerning loss function. The gradient is now projected orthogonally at step 15 before backpropagation (described in section~\ref{sec:train_known_task}). Eventually, the buffer memory is reorganized at step 16 of Algorithm~\ref{alg:training}.
\begin{algorithm}[!htb]
    \caption{Open world labeling}
    \label{alg:owl-labeling}
    \begin{algorithmic}[1]
        \STATE {\bfseries Inputs:} fully unlabeled dataset of task `$t$' is $\mathcal{D}^{t} = \{\mathcal{D}_{u}^{t}\}$, 
        agreement fraction for attack and benign classes $\rho_{a}, \rho_{b}$,
         expected class imbalance ratio $\pi$ (computed by taking the average over previously seen tasks), 
         minimum expected confidence for attack and benign classes $\gamma_a, \gamma_b$,
         labeled data ratio $r$, and 
         buffer memory $\mathcal{M}$, $f^{t-1}_{\theta}$ - model trained until task `$t-1$'.
        
        \STATE {\bfseries Output:} Modified dataset for task `$t$' with labeled and unlabeled portions $\mathcal{\hat{D}}^{t} = \{\mathcal{\hat{D}}_{l}^{t}, \mathcal{\hat{D}}_{u}^{t}\}$.
        
        \STATE $\mathcal{C}_{a} \leftarrow$ probability values (softmax confidences) corresponding to the attack class for samples in $\mathcal{D}^{t}_{u}$ where $f^{t-1}_{\theta}$ predicted as an attack 
        \STATE $\mathcal{C}_{b} \leftarrow$ probability values (softmax confidences) corresponding to the benign class for samples in $\mathcal{D}^{t}_{u}$ where $f^{t-1}_{\theta}$ predicted as a benign

        \STATE $\{X^{top}_{a}\} \leftarrow$ set of samples belong to the attack class (pseudo labeled by $f^{t-1}_{\theta}$) for which $C_{a} > \gamma_{a}$
        \STATE $\{X^{top}_{b} \}\leftarrow$ set of samples belong to benign class (pseudo labeled by $f^{t-1}_{\theta}$) for which $C_{b} > \gamma_{b}$
        
        \STATE initialise $\mathcal{S}_a = \mathcal{S}_b = $ \{\} 
        (set of mutually agreed-upon samples for attack and benign classes)

        \FOR {$\mathbf{x}$ in $X^{top}_{a}$}
            \STATE $\{m_\mathbf{x} \}\leftarrow$ set of samples in $\mathcal{M}$ closest to $\mathbf{x}$ (computed using cosine distance)
            \STATE $\mathbf{y}_\mathbf{x} =$ majority of set of labels assigned to $ \{ m_\mathbf{x}\}$
            \IF{$\mathbf{y}_\mathbf{x} ==$ `attack'}
                \STATE add $\{\mathbf{x}, $`attack'$\}$ to $\mathcal{S}_a$
            \ENDIF
        \ENDFOR
        
        \FOR {$\mathbf{x}$ in $X^{top}_{b}$}
            \STATE $\{m_\mathbf{x} \}\leftarrow$ set of samples in $\mathcal{M}$ closest to $\mathbf{x}$ (computed using cosine distance)
            \STATE $\mathbf{y}_\mathbf{x} =$ majority of set of labels assigned to $ \{ m_\mathbf{x}\}$
            \IF{$\mathbf{y}_\mathbf{x} ==$ `benign'}
                \STATE add $\{\mathbf{x}, $`benign'$\}$ to $\mathcal{S}_b$
            \ENDIF
        \ENDFOR

        \STATE $\mathcal{\hat{D}}^{t}_{l} = \{ \rho_a \times |\mathcal{S}_a|$ samples from $\mathcal{S}_a\}$ $\cup$ $\{ \rho_b \times |\mathcal{S}_b|$ samples from $\mathcal{S}_b\}$
        
        \STATE Estimated number of attack and benign samples \\ $n_{a} = \pi \times|\mathcal{D}^{t}|$,  
        $n_{b} = (1 - \pi)\times|\mathcal{D}^{t}|$ 

        \STATE Remaining samples to be labeled by the security analyst
        \\ $e_{a} = r \times n_{a} - \rho_a \times |\mathcal{S}_a|$ 
        \\ $e_{b} = r \times n_{b} - \rho_b \times |\mathcal{S}_b|$ 

        \STATE Get $e_a$ attack samples and $e_b$ benign samples from the security analyst and add them to $\mathcal{\hat{D}}^{t}_{l}$ 

        \STATE $\mathcal{\hat{D}}^{t}_{u}$ = $\mathcal{D}^{t} - \mathcal{\hat{D}}^{t}_{l}$

        \RETURN $\{\mathcal{\hat{D}}^{t}_{l}, \mathcal{\hat{D}}^{t}_{u}\}$

    \end{algorithmic}
  
\end{algorithm}
\par For the unseen tasks, Algorithm~\ref{alg:training} at step 13 invokes the open world labeling algorithm (Algorithm~\ref{alg:owl-labeling}), that generates the labels using the procedure described in Section~\ref{sec:train_unseen}. Using the labeled data, Algorithm~\ref{alg:training} at step 14 invokes the semi-supervised continual learning algorithm (Algorithm~\ref{alg:semi-supervised}), and buffer memory is reorganized at step 16 of Algorithm~\ref{alg:training}.

 \section{Experiments and Analysis}
  \label{sec:results}
  \textbf{Datasets:} We use four NID datasets in our experiments. Specifically, CTU-13,~\cite{garcia2014empirical}, CICIDS-2017~\cite{liu2022error}, CSE-CICIDS-2018~\cite{liu2022error}, and UNSW-NB15~\cite{7348942}. 
  The datasets of the CICIDS-2017~\cite{liu2022error} and 2018~\cite{liu2022error} are spread over multiple files in comma-separated value (CSV) format, each containing 90 features. For each CSV file, flow label identifiers (\texttt{flowid, source/destination IP address, and source/destination port}) and duplicate rows are removed, whereas not a number (NaN) values are replaced with the corresponding feature mean value. Eventually, min-max normalization is applied for each column using the \textit{sklearn} library. All the attack class labels with suffix \textit{attempted} are relabeled as benign labels.  The preprocessed dataset of the CTU-13 dataset available in \cite{amalapuram2023augmented} is used in our work, and it contains a total of 39 features. For the UNSW-NB15 dataset, the features \texttt{srcip, sport, dstip, dsport, Stime, and Ltime} features are removed. Later, we perform a categorical data encoding for the features like \texttt{proto, state, and service}, followed by min-max normalization, after which the total number of features is 202.

  \par \textbf{Task formulation:} For CICIDS 2017 and CSE-CICIDS-2018 datasets, the granularity of the task creation is a single day as their traffic spreads over five and ten days, respectively, so we created five and ten tasks respectively. Similar to ~\cite{amalapuram2023augmented}, we created five tasks for CTU-13. For UNSW-NB15, as it contains only two days of traffic, we created nine tasks as the dataset contains nine different attacks, and the entire benign class data is also split and shared across the nine tasks.

  \par \textbf{Baseline selection:} 
  We consider regularization, projection-based, memory-replay-based, and class-imbalanced-centered methods from supervised CL methods. These methods include elastic weight consolidation (EWC~\cite{kirkpatrick2017overcoming}), synaptic intelligence (SI~\cite{zenke2017continual}), gradient episodic memory (GEM~\cite{lopez2017gradient}), average-gradient episodic memory (A-GEM~\cite{chaudhry2018efficient}), maximal interfered retrieval (MIR~\cite{aljundi2019online}), class-balanced reservoir sampling (CBRS~\cite{chrysakis2020online}). These methods are trained under full supervision. 
   Most of the semi-supervised CL methods~\cite{wang2021ordisco,smith2021memory,berthelot2019mixmatch,sohn2020fixmatch,kang2023soft} proposed in the literature use label preserving data augmentation (DA) techniques like cropping, rotating, and artificial noise injection for validating consistency regularization while generating pseudo labels for unlabeled data. Further, adopting the existing methods by eliminating DA methods may result in inconsistent evaluation. So, we don't consider SSCL techniques in our baseline methods. \textcolor{black}{To solve this issue, we partly followed the evaluation protocol defined in~\cite{apruzzese2022sok} under a semi-supervised setting. We compare the performance of the proposed method (that works with limited data) with baselines that operate in a fully supervised fashion. This comparison helps understand the gap between the performance of supervised methods and that of the proposed method.} Prior work~\cite{amalapuram2023augmented} have already shown the ineffectiveness of the shallow methods (random forest) in the CL framework, so this work also don't consider them for baseline.  
  
\par \textbf{Evaluation metrics}: 
In our experiments, we use the precision-recall area under the curve (PR-AUC) to measure the detection performance of both benign and attack traffic (malicious apps). We use the area under the time (AUT) performance metric~\cite{pendlebury2019tesseract} to evaluate the proposed method's effectiveness. In other words, AUT quantifies the performance degradation of the model over time. AUT is computed over the base metrics like PR-AUC as shown in Equation~\ref{eq:aut},
\begin{equation} \label{eq:aut}
\begin{split}
AUT(f,N) = \frac{1}{N-1}\sum_{k=1}^{N-1}\frac{[f(x_{k+1})+f(x_{k})]}{2},
\end{split}
\end{equation}
where $f(x_{k})$ is the point estimate computed using the performance metric $f$ at the time period $k$, $N$ is the number of time slots used to test performance degradation, and $1/(N-1)$ is the normalization constant.

\par \textbf{Experimental design}: The details of the unseen tasks per each dataset are as follows (total number of tasks/seen tasks/unseen tasks); CTU-13 (5/2/3), CICIDS-2017 (5/2/3), CSE-CICIDS-2018 (10/5/5), and UNSW-NB15 (9/3/7). All the baselines are 
 trained in a fully supervised continual learning fashion on all tasks. SOUL trains on seen tasks following the procedure in Section~\ref{sec:train_known_task}. Likewise, it utilizes the procedure from Section~\ref{sec:train_unseen} for unseen tasks. Later, PR-AUC values for the benign and attack classes are computed on all the tasks. Eventually, the AUT values are computed for all PR-AUC values. We categorized the evaluation as follows:
\begin{itemize}
    \item \textit{Evaluation of SOUL-seen tasks} related to performance results (seen-AUT (A), seen-AUT (B)) on the tasks whose partial labels are already present.
    \item \textit{Evaluation of SOUL's unseen tasks} related to performance results (unseen-AUT (A), unseen-AUT (B)) on the unseen tasks whose partial labels are derived from the classifier and analysts (novel attacks).
    \item \textit{Evaluation of overall tasks} of SOUL related to performance results (overall-AUT (A), overall-AUT (B)) on the seen and unseen tasks.
    \item \textit{All the baselines} (irrespective of seen, unseen, and overall tasks) are trained in a fully supervised fashion.
\end{itemize}
\begin{table*}
  \caption{Comparing the performance results of the proposed SOUL method with baselines on CTU-13, UNSW-NB15, CICIDS-2017, and CICIDS-2018 datasets. We report AUT values for benign and attack classes on seen and unseen tasks and overall tasks. All the baseline methods are trained on unseen tasks in a fully supervised setting, and best values are marked in \textbf{bold}.}
  \label{tab:soul_comparison_with_unseen}
  \centering
  \begin{tabular}{llllllllllllllllllllllllllllll}
    \toprule   
    & \multicolumn{6}{c}{CTU-13} \\
    \cmidrule(lr){1-2} \cmidrule(lr){2-7}
   Baseline Methods & seen-AUT (B) & seen-AUT (A) & unseen-AUT (B) & unseen-AUT (A) & overall-AUT (B) & overall-AUT (A)\\
   \midrule
  
   EWC~\cite{kirkpatrick2017overcoming} & 0.999 $\pm$ 0.000 & 0.008 $\pm$ 0.002 & 0.963 $\pm$ 0.001 & 0.590 $\pm$ 0.053 & 0.980 $\pm$ 0.000 & 0.389 $\pm$ 0.047 \\
 GEM~\cite{lopez2017gradient} & 0.999 $\pm$ 0.000 & 0.119 $\pm$ 0.035 & 0.939 $\pm$ 0.003 & 0.648 $\pm$ 0.074 & 0.964 $\pm$ 0.002 & 0.417 $\pm$ 0.067 \\
 AGEM~\cite{chaudhry2018efficient} & 0.999 $\pm$ 0.000 & 0.077 $\pm$ 0.044 & 0.940 $\pm$ 0.003 & 0.602 $\pm$ 0.078 & 0.964 $\pm$ 0.002 & 0.371 $\pm$ 0.076 \\
 SI~\cite{zenke2017continual} & 0.999 $\pm$ 0.000 & 0.079 $\pm$ 0.046 & 0.940 $\pm$ 0.004 & 0.606 $\pm$ 0.083& 0.964 $\pm$ 0.002 & 0.374 $\pm$ 0.079\\
 CBRS~\cite{chrysakis2020online} & 0.999 $\pm$ 0.000 & 0.985 $\pm$ 0.007 & \textbf{0.997 $\pm$ 0.002} & 0.993 $\pm$ 0.007 & 0.998 $\pm$ 0.001 & 0.991 $\pm$ 0.005\\
 MIR~\cite{aljundi2019online} & 0.999 $\pm$ 0.000 & 0.993 $\pm$ 0.000 & 0.996 $\pm$ 0.004 & \textbf{0.996 $\pm$ 0.003} & \textbf{0.998 $\pm$ 0.002} & 0.994 $\pm$ 0.001\\
 \midrule
 SOUL & \textbf{0.999 $\pm$ 0.000} &\textbf{ 0.997 $\pm$ 0.001} & 0.994 $\pm$ 0.003 & 0.993 $\pm$ 0.004 & 0.997 $\pm$ 0.001 & \textbf{0.994 $\pm$ 0.003}\\
    
    \bottomrule
  \end{tabular}   

  \begin{tabular}{llllllllllllllllllllllllllllll}
        & \multicolumn{6}{c}{UNSW-NB15} \\
    \cmidrule(lr){1-2} \cmidrule(lr){2-7}
   Baseline Methods & seen-AUT (B) & seen-AUT (A) & unseen-AUT (B) & unseen-AUT (A) & overall-AUT (B) & overall-AUT (A)\\
   \midrule
   EWC &  0.999 $\pm$ 0.000 & 0.424 $\pm$ 0.157 & 0.998 $\pm$ 0.000 & 0.663 $\pm$ 0.036 & 0.998 $\pm$ 0.000 & 0.585 $\pm$ 0.086\\  
 GEM &  0.983 $\pm$ 0.003 & 0.436 $\pm$ 0.305 & 0.898 $\pm$ 0.017 & 0.296 $\pm$ 0.130 & 0.930 $\pm$ 0.012 & 0.334 $\pm$ 0.185\\
 AGEM & 0.983 $\pm$ 0.000 & 0.029 $\pm$ 0.001 & 0.898 $\pm$ 0.000 &  0.130 $\pm$ 0.008 & 0.930 $\pm$ 0.000 & 0.093 $\pm$ 0.007 \\
 SI & 0.982 $\pm$ 0.000 & 0.022 $\pm$ 0.001 & 0.897 $\pm$ 0.000 & 0.124 $\pm$ 0.004 & 0.928 $\pm$ 0.000  & 0.086 $\pm$ 0.001\\
 CBRS &  0.996 $\pm$ 0.004 & 0.333 $\pm$ 0.131 & 0.982 $\pm$ 0.002 & 0.539 $\pm$ 0.012 & 0.987 $\pm$ 0.013 & 0.463 $\pm$ 0.054 \\
 MIR & 0.999 $\pm$ 0.001 & 0.838 $\pm$ 0.025 & 0.999 $\pm$ 0.001 & \textbf{0.795 $\pm$ 0.012} & 0.999 $\pm$ 0.001 & \textbf{0.823 $\pm$ 0.001}\\
 \midrule
 SOUL & \textbf{0.999 $\pm$ 0.000} & \textbf{0.845 $\pm$ 0.012} & \textbf{0.999 $\pm$ 0.000} & 0.727 $\pm$ 0.018 & 0\textbf{.999 $\pm$ 0.000}& 0.783 $\pm$ 0.014\\

    \bottomrule
  \end{tabular}   

     \begin{tabular}{llllllllllllllllllllllllllllll}
    \toprule

    & \multicolumn{6}{c}{CICIDS-2017} \\
    \cmidrule(lr){1-2} \cmidrule(lr){2-7}
   Baseline Methods & seen-AUT (B) & seen-AUT (A) & unseen-AUT (B) & unseen-AUT (A) & overall-AUT (B) & overall-AUT (A)\\
   \midrule
   EWC & 0.993 $\pm$ 0.003 & 0.042 $\pm$ 0.027 & 0.968 $\pm$ 0.019 & 0.747 $\pm$ 0.108 & 0.969 $\pm$ 0.018 & 0.455 $\pm$ 0.097\\
   GEM & 0.988 $\pm$ 0.000 & 0.016 $\pm$ 0.000 & 0.811 $\pm$ 0.004 & 0.216 $\pm$ 0.006 & 0.874 $\pm$ 0.002 & 0.144 $\pm$ 0.003 \\
   AGEM & 0.989 $\pm$ 0.000 & 0.014 $\pm$ 0.000 & 0.807 $\pm$ 0.001 & 0.206 $\pm$ 0.004 & 0.872 $\pm$ 0.001 & 0.137 $\pm$ 0.002 \\
   SI & 0.989 $\pm$ 0.000 & 0.014 $\pm$ 0.000 & 0.807 $\pm$ 0.001 & 0.206 $\pm$ 0.004 & 0.872 $\pm$ 0.000 & 0.137 $\pm$ 0.002 \\
   CBRS & 0.999 $\pm$ 0.000 & 0.461 $\pm$ 0.139 & 0.996 $\pm$ 0.001 & 0.940 $\pm$ 0.011 & 0.997 $\pm$ 0.000 & 0.778 $\pm$ 0.057 \\
   MIR & 0.998 $\pm$ 0.000 & 0.093 $\pm$ 0.034 & 0.997 $\pm$ 0.000 & 0.947 $\pm$ 0.016& 0.996 $\pm$ 0.000 & 0.618 $\pm$ 0.030 \\
   \midrule
   SOUL & \textbf{0.999 $\pm$ 0.000} & \textbf{0.995 $\pm$ 0.003} & \textbf{0.998 $\pm$ 0.000} & \textbf{0.982 $\pm$ 0.007} & 0\textbf{.999 $\pm$ 0.000} & \textbf{0.989 $\pm$ 0.003} \\
    
    \bottomrule
  \end{tabular}  
     \begin{tabular}{llllllllllllllllllllllllllllll}
    \toprule       
    & \multicolumn{6}{c}{CSE-CICIDS-2018}\\
    \cmidrule(lr){1-2} \cmidrule(lr){2-7} 
   Baseline Methods & seen-AUT (B) & seen-AUT (A) & unseen-AUT (B) & unseen-AUT (A) & overall-AUT (B) & overall-AUT (A)\\
   \midrule
   EWC  & 0.897 $\pm$ 0.057 & 0.086 $\pm$ 0.045 & 0.985 $\pm$ 0.009 & 0.039 $\pm$ 0.022 & 0.958 $\pm$ 0.023 & 0.059 $\pm$ 0.036\\ 
   GEM &  0.907 $\pm$ 0.002 & 0.113 $\pm$ 0.022 & 0.986 $\pm$ 0.000 & 0.019 $\pm$ 0.007 & 0.961 $\pm$ 0.001 & 0.053 $\pm$ 0.017\\
   AGEM &  0.921 $\pm$ 0.004 & 0.093 $\pm$ 0.000 & 0.988 $\pm$ 0.000 & 0.012 $\pm$ 0.000 & 0.967$\pm$ 0.001 & 0.036 $\pm$ 0.000\\
   SI &  0.924 $\pm$ 0.002 & 0.093 $\pm$ 0.001 & 0.988 $\pm$ 0.000 & 0.012 $\pm$ 0.000 & 0.968 $\pm$ 0.000 & 0.037 $\pm$ 0.001\\
   CBRS &  0.954 $\pm$ 0.000 & 0.545 $\pm$ 0.000 & 0.994 $\pm$ 0.000 & 0.505 $\pm$ 0.000 & 0.982 $\pm$ 0.001 &  0.517 $\pm$ 0.000\\
   MIR &  0.954 $\pm$ 0.000 & 0.545 $\pm$ 0.000 & 0.994 $\pm$ 0.000 & 0.505 $\pm$ 0.000 & 0.982 $\pm$ 0.001 & 0.517 $\pm$ 0.000\\
   \midrule
   SOUL & \textbf{0.999 $\pm$ 0.000} & \textbf{0.761 $\pm$ 0.016} & \textbf{0.999 $\pm$ 0.000} & \textbf{0.725 $\pm$ 0.100} & \textbf{0.999 $\pm$ 0.000} & \textbf{0.675 $\pm$ 0.065}\\   
    
    \midrule
  \end{tabular}  
\end{table*}
\par \textbf{Quantitative analysis}: The performance results of the proposed SOUL method are present in Table~\ref{tab:soul_comparison_with_unseen}. We make the following observations. First, the proposed SOUL method outperforms the baselines on the CICIDS-2017,2018 datasets and ranks second on the UNSW-NB15 and CTU-13 datasets. However, the difference on the performance comparison on CTU-13 is subtle, as it only appears in the third decimal place.   
\begin{table}[tbh!]
  \caption{Comparing the number of labels generated by the model and the security analyst for the unseen tasks. The \%savings indicates the percentage of the labels generated by the model out of the total labeled samples (Analyst + Model).}
  \label{tab:labels_saving}
  \centering
 
  \begin{tabular}{llllllllllllllllllllllllllllll}
      \cmidrule(lr){1-5} 
   Dataset & \#Model & \#Analyst& \#Total & \%savings \\
   \midrule
   
   CTU-13 & 62906 & 72101 & 135007 & 46.5\%\\
   UNSW-NB15 & 28867 & 224461 & 253328 & 11.3\%\\
   CICIDS-2017  &  35093& 167899 & 202992& 17.2\% \\
CSE-CICIDS-2018 &  1067780 & 2668975 & 3736755 & 28.5\%\\  
    
    \bottomrule    
  \end{tabular}  
\vspace{-6mm}
\end{table}
Second, the performance of the baselines CBRS and MIR are consistent across all the datasets; this can be attributed to their inherent sample replay mechanism from memory, which handles CI and CF effectively. Third, the remaining baselines exhibit CF on initial tasks, as evidenced by their poor performance on seen-AUT(A) values. Intuitively, this may be due to the SGD optimizer yielding lower error on the most recent tasks, as indicated by higher AUT values on unseen-AUT (A and B) across all baseline methods. These methods lack an additional orthogonal projection component to preserve past knowledge. This empirical result aligns with the motivation for incorporating orthogonal projections as an additional inductive bias, as described in the Section~\ref{subsec:work-motivation}. Eventually, the efficacy of SOUL is also evaluated on the number of labels generated for the unseen test samples. From Table~\ref{tab:labels_saving}, we observe that the proposed method generated labels are in the range of 11 to 46\% relative to the total number of labeled samples (of unseen tasks). 

\label{mean_cir_effectiness}
\par \textbf{Impact of the estimated CIR:} 
In this section, we empirically verify the effectiveness of the mean CIR on the detection performance of the unseen tasks. Let us define CIR as the ratio of the total number of attack samples to the total number of samples present in a given task. When using mean CIR as an expected CIR of unseen tasks, two possible situations can arise here:
\begin{enumerate}
    \item \textcolor{black}{If the actual CIR of the unseen task is lower than the expected CIR, the SOUL method generates labels for novel samples by combining classifier confidence, buffer memory, and analyst inputs.}
\item \textcolor{black}{If the actual CIR of the unseen task exceeds the expected CIR, SOUL generates fewer labels for novel samples than anticipated in a semi-supervised setting. Despite this, SOUL's performance remains comparable to that of
fully supervised baselines.} We demonstrate how the SOUL method effectively handles the second situation using the CICIDS-2017 dataset. Table~\ref{tab:CIR} details the sample count, CIR, and expected labeled samples per task. For this experiment, we set the labeled data proportion to 20\% per task.
\end{enumerate}

\begin{table*}[!tbh]
  \caption{\textcolor{black}{Presents the number of samples, class imbalance ratio (CIR), and the expected number of unlabeled
samples per task for the CICIDS-2017 dataset. The table also specifies the labeled data proportion (20\% for all
tasks) and the meaning of `B' (benign) and `A' (attack) in the fine-grained sample rows.}}
  \label{tab:CIR}
  \centering
 
 \begin{adjustbox}{width=1\textwidth}
  \begin{tabular}{llllllllllllllllllllllllllllll}      
    & \multicolumn{2}{c}{Seen tasks} & \multicolumn{3}{c}{Unseen tasks} \\
    \cmidrule(lr){2-3} \cmidrule(lr){4-6} 
  Task id & Task 1 & Task 2  &  Task 3 & Task 4 & Task 5  \\ \hline
  samples in the task & 229737 & 270453 & 354613 & 260670 & 399692 \\ \hline
  Fine granularity samples & \parbox{3cm}{`B':224540 `A':5197} & \parbox{3cm}{`B':264779 `A':5674} & \parbox{3cm}{`B':231558 `A':123055} & \parbox{3cm}{`B':206744 `A':53926} & \parbox{3cm}{`B':208485 `A':191207} \\ \midrule
  CIR & 0.0226215 & 0.02097961 & 0.3470120 & 0.206874 & 0.4783858\\ \hline
  \parbox{5cm}{Expected no.of unseen task samples (20\% labeled data) to get labels per
task} & -&  -&  70922 &52134 & 79938\\ \hline
\parbox{5cm}{Expected no of labeled samples (20\% labeled data) at fine-granularity level} & - & - & \parbox{3cm}{`B':46311 `A':24611} & \parbox{3cm}{`B':41348 `A':10785} & \parbox{3cm}{`B':41697 `A':38241} \\ \hline

  \end{tabular}
  
  \end{adjustbox}
  
  \end{table*}

  Following training on the first two tasks, the expected CIR for unseen tasks was calculated as 0.021800555.
However, this estimate was lower than the actual CIR values of the unseen tasks. Consequently, the actual number
of labeled attack class samples for unseen tasks was reduced, as detailed in Table~\ref{tab:CIR_2}.

\begin{table*}[!tbh]
  \caption{\textcolor{black}{Illustrating the discrepancy between the expected (as per actual CIR) and actual (as per expected CIR) number of labeled samples for unseen tasks, the table shows the number of samples labeled by the classifier and
analyst based on the expected CIR for the CICIDS-2017 dataset.}}
  \label{tab:CIR_2}
  \centering
 
  \begin{tabular}{llll}      
    & \multicolumn{3}{c}{Unseen tasks}  \\
    \cmidrule(lr){2-4} 
  Task id &  Task 3 & Task 4 & Task 5  \\ \hline
  
  \parbox{7cm}{Expected no of labeled samples (20\% labeled data) at fine-
granularity level (as per actual CIR)} & \parbox{3cm}{`B':46311 `A':24611} & \parbox{3cm}{`B':41348 `A':10785} & \parbox{3cm}{`B':41697 `A':38241} \\ \hline
 
\parbox{7cm}{Actual no of labeled samples (20\% labeled data) at fine-granularity
level (as per expected CIR)} & \parbox{3cm}{`B':69377 `A':1545} & 
\parbox{3cm}{`B':50998 `A':1136} & \parbox{3cm}{`B':78196 `A':1741} \\ \hline

  \end{tabular}
  
  
  \end{table*}
Despite having fewer labeled attack samples for unseen tasks, SOUL's performance remains comparable to that of
fully supervised baselines. Additionally, we compared SOUL's performance to a scenario where all expected labels for unseen tasks (based on the actual CIR) were provided solely by analysts. Table~\ref{tab:comparing_label} presents these results. Notably, SOUL's performance remained comparable to baseline models even under these conditions. Thus, the SOUL method effectively handles scenarios where the expected CIR deviates from the actual CIR
of unseen tasks.

\begin{table*}
  \caption{We compared the performance of the proposed SOUL methods using two label generation approaches for
unseen tasks: one relying on both classifier and analyst (C+A) for labeling as per the expected CIR, the other solely on analyst (A) for labeling as per the actual CIR. Experiments were conducted on CICIDS-2017, UNSW-NB15, CTU-13, and CICIDS-2018
datasets and three different seed values and mean values were reported.}
  \label{tab:comparing_label}
  \centering
 
  \begin{tabular}{llllllllllllllllllllllllllllll}      
    \cmidrule(lr){3-8}  
  Dataset & \parbox{1.9cm}{Label generation process} & seen-AUT (B) & seen-AUT (A) & unseen-AUT (B) & unseen-AUT (A) & overall-AUT (B) & overall-AUT (A)\\
   \midrule
   CTU13 & C+A & 0.999 $\pm$ 0.000 & 0.997 $\pm$ 0.001 & 0.994 $\pm$ 0.003 & 0.993 $\pm$ 0.004 & 0.997 $\pm$ 0.001 & 0.994 $\pm$ 0.003\\
  & A  & 0.999 $\pm$ 0.000 & 0.996 $\pm$ 0.000 & 0.995 $\pm$ 0.004 & 0.996 $\pm$ 0.002 & 0.997 $\pm$ 0.002 & 0.996 $\pm$ 0.001 \\
   \midrule
   UNSW-NB15 & C+A & 0.999 $\pm$ 0.000 & 0.845 $\pm$ 0.012 & 0.999 $\pm$ 0.000 & 0.727 $\pm$ 0.018 & 0.999 $\pm$ 0.000& 0.783 $\pm$ 0.014\\
   & A & 0.999 $\pm$ 0.000 & 0.836 $\pm$ 0.010 & 0.999 $\pm$ 0.000 & 0.740 $\pm$ 0.018 & 0.999 $\pm$ 0.000 & 0.788 $\pm$ 0.014\\
   \midrule
   CICIDS-2017 & C+A & 0.999 $\pm$ 0.000 & 0.995 $\pm$ 0.003 & 0.998 $\pm$ 0.000 & 0.982 $\pm$ 0.007 & 0.999 $\pm$ 0.000 & 0.989 $\pm$ 0.003\\
   & A & 0.999 $\pm$ 0.000 & 0.956 $\pm$ 0.012 & 0.999 $\pm$ 0.000 & 0.986 $\pm$ 0.004 & 0.999 $\pm$ 0.000 & 0.981 $\pm$ 0.002 \\
   \midrule
   CSE-CICIDS-2018 & C+A & 0.999 $\pm$ 0.000 & 0.761 $\pm$ 0.016 & 0.999 $\pm$ 0.000 & 0.725 $\pm$ 0.100 & 0.999 $\pm$ 0.000 & 0.675 $\pm$ 0.065\\
     & A & 0.975 $\pm$ 0.017 & 0.638 $\pm$ 0.143 & 0.998 $\pm$ 0.001 & 0.584 $\pm$ 0.115 & 0.988 $\pm$ 0.008 & 0.582 $\pm$ 0.092\\
     \bottomrule
  \end{tabular}

  \end{table*}

\subsection{Ablation studies} 
\label{sec:ablation-study}
\par \textbf{Sensitivity of each component:} We investigate the robustness of the proposed SOUL’s performance by examining its sensitivity to buffer memory and gradient projection memory. From Table~\ref{tab:ablation_study}, we observe that each component consistently impacts the proposed method's performance across all the benchmark datasets. 
The empirical evaluation reveals that using buffer memory and orthogonal projections offers significant advantages for the proposed method's performance.

\begin{table*}[!tbh]
  \caption{Ablation study demonstrating the sensitivity of the SOUL method various components including buffer memory and gradient projections memory. The best values are marked in \textbf{bold}.}
  \label{tab:ablation_study}
  \centering
 
  

   \begin{tabular}{llllllll}
   \toprule
  \multicolumn{8}{c}{CTU-13}\\
      \cmidrule(lr){3-8} 
   Memory & GPM &  seen-AUT (B) & seen-AUT (A) & unseen-AUT (B) & unseen-AUT (A) & overall-AUT (B) & overall-AUT (A)\\
   \midrule
   
    \nomark & \nomark & 0.999 $\pm$ 0.000 & 0.738 $\pm$ 0.321 & 0.957 $\pm$ 0.009 & 0.877 $\pm$ 0.040 & 0.975 $\pm$ 0.005 & 0.801 $\pm$ 0.155\\
    \yesmark & \nomark & 0.999 $\pm$ 0.000 & 0.738 $\pm$ 0.322 & 0.958 $\pm$ 0.009 & 0.876 $\pm$ 0.048 & 0.975 $\pm$ 0.005 & 0.801 $\pm$ 0.016\\
      \nomark& \yesmark & 0.999 $\pm$ 0.000 & 0.900 $\pm$ 0.119 & 0.975 $\pm$ 0.006 & 0.928 $\pm$ 0.018 & 0.986 $\pm$ 0.004 & 0.905 $\pm$ 0.059\\
      \midrule
      \yesmark & \yesmark & \textbf{0.999 $\pm$ 0.000} & \textbf{0.997 $\pm$ 0.001} & \textbf{0.994 $\pm$ 0.003} & \textbf{0.993 $\pm$ 0.004} & \textbf{0.997 $\pm$ 0.001} & \textbf{0.994 $\pm$ 0.003}\\  
    
    \bottomrule    
  \end{tabular}

  \begin{tabular}{llllllll}
   \toprule
  \multicolumn{8}{c}{UNSW-NB15}\\
      \cmidrule(lr){3-8} 
   Memory & GPM &  seen-AUT (B) & seen-AUT (A) & unseen-AUT (B) & unseen-AUT (A) & overall-AUT (B) & overall-AUT (A)\\
   \midrule
   
    \nomark & \nomark & 0.999 $\pm$ 0.000 & 0.668 $\pm$ 0.101 & 0.998 $\pm$ 0.000 & 0.668 $\pm$ 0.003 & 0.998 $\pm$ 0.000 & 0.690 $\pm$ 0.057\\
    \yesmark & \nomark & 0.999 $\pm$ 0.000 & 0.668 $\pm$ 0.095 & 0.998 $\pm$ 0.000 & 0.663 $\pm$ 0.025 & 0.998 $\pm$ 0.000 & 0.679 $\pm$ 0.052\\
      \nomark& \yesmark & 0.999 $\pm$ 0.000 & 0.658 $\pm$ 0.145 & 0.998 $\pm$ 0.000 & 0.724 $\pm$ 0.011 & 0.999 $\pm$ 0.000 & 0.711 $\pm$ 0.065\\
      \midrule
      \yesmark & \yesmark & \textbf{0.999 $\pm$ 0.000} & \textbf{0.845 $\pm$ 0.012} & \textbf{0.999 $\pm$ 0.000} & \textbf{0.727 $\pm$ 0.018} & \textbf{0.999 $\pm$ 0.004}& \textbf{0.783 $\pm$ 0.014}\\  
    
    \bottomrule    
  \end{tabular}

  \begin{tabular}{llllllll}
   \toprule
  \multicolumn{8}{c}{CICIDS-2017}\\
      \cmidrule(lr){3-8} 
   Memory & GPM &  seen-AUT (B) & seen-AUT (A) & unseen-AUT (B) & unseen-AUT (A) & overall-AUT (B) & overall-AUT (A)\\
   \midrule
   
    \nomark & \nomark & 0.996 $\pm$ 0.001 & 0.408 $\pm$ 0.050 & 0.881 $\pm$ 0.080 & 0.348 $\pm$ 0.087 & 0.923 $\pm$ 0.043 & 0.313 $\pm$ 0.044\\
    \yesmark & \nomark & 0.997 $\pm$ 0.000 & 0.359 $\pm$ 0.105 & 0.886 $\pm$ 0.082 & 0.338 $\pm$ 0.069 & 0.928 $\pm$ 0.043 & 0.299 $\pm$ 0.039\\
      \nomark& \yesmark & 0.996 $\pm$ 0.000 & 0.361 $\pm$ 0.050 & 0.995 $\pm$ 0.001 & 0.933 $\pm$ 0.031 & 0.996 $\pm$ 0.000 & 0.757 $\pm$ 0.037\\
      \midrule
      \yesmark & \yesmark & \textbf{0.999 $\pm$ 0.000} & \textbf{0.995 $\pm$ 0.003} & \textbf{0.998 $\pm$ 0.000} & \textbf{0.982 $\pm$ 0.007} & \textbf{0.999 $\pm$ 0.000} & \textbf{0.989 $\pm$ 0.003}\\  
    
    \bottomrule    
  \end{tabular}

  \begin{tabular}{llllllll}
   \toprule
  \multicolumn{8}{c}{CSE-CICIDS-2018}\\
      \cmidrule(lr){3-8} 
   Memory & GPM &  seen-AUT (B) & seen-AUT (A) & unseen-AUT (B) & unseen-AUT (A) & overall-AUT (B) & overall-AUT (A)\\
   \midrule
   
    \nomark & \nomark & 0.971 $\pm$ 0.017 & 0.370 $\pm$ 0.212 & 0.995 $\pm$ 0.001 & 0.052 $\pm$ 0.038 & 0.985 $\pm$ 0.007 & 0.196 $\pm$ 0.093\\
    \yesmark & \nomark & 0.960 $\pm$ 0.003 & 0.425 $\pm$ 0.157 & 0.996 $\pm$ 0.000 & 0.401 $\pm$ 0.143 & 0.981 $\pm$ 0.001 & 0.416 $\pm$ 0.142\\
      \nomark& \yesmark & 0.976 $\pm$ 0.017 & 0.327 $\pm$ 0.222 & 0.996 $\pm$ 0.001 & 0.245 $\pm$ 0.153 & 0.988 $\pm$ 0.007 & 0.254 $\pm$ 0.058\\
      \midrule
      \yesmark & \yesmark & \textbf{0.999 $\pm$ 0.000} & \textbf{0.761 $\pm$ 0.016} & \textbf{0.999 $\pm$ 0.000} & \textbf{0.725 $\pm$ 0.100} & \textbf{0.999 $\pm$ 0.000} & \textbf{0.675 $\pm$ 0.065}\\  
    
    \bottomrule    
  \end{tabular}
 
   \end{table*}

\begin{table*}[!tbh]
  \caption{Demonstrating the effect of varying the amount of labeled data on the performance results of the proposed SOUL method. The increase in the result value on a particular metric is marked using $\uparrow$ and decrease with $\downarrow$.}
  \label{tab:ablation_study_varying_label_ratio}
  \centering
 
 \begin{tabular}{llllllll}
   \toprule
  \multicolumn{7}{c}{CTU-13}\\
      \cmidrule(lr){2-7} 
   Labeled data ratio &  seen-AUT (B) & seen-AUT (A) & unseen-AUT (B) & unseen-AUT (A) & overall-AUT (B) & overall-AUT (A)\\
   \midrule
   
    2\% & 0.999 $\pm$ 0.000 & 0.989 $\pm$ 0.003 & 0.985 $\pm$ 0.007 & 0.966 $\pm$ 0.019 & 0.991 $\pm$ 0.003 & 0.969 $\pm$ 0.014\\
    10\% & 0.999 $\pm$ 0.000 & 0.997 $\pm$ 0.000 ($\uparrow$) & 0.990 $\pm$ 0.003 ($\uparrow$) & 0.992 $\pm$ 0.001 ($\uparrow$) & 0.995 $\pm$ 0.001 ($\uparrow$)& 0.993 $\pm$ 0.000 ($\uparrow$)\\      
      
      20\% & 0.999 $\pm$ 0.000 & 0.997 $\pm$ 0.001 & 0.994 $\pm$ 0.003 ($\uparrow$)& 0.993 $\pm$ 0.004 ($\uparrow$)& 0.997 $\pm$ 0.001 ($\uparrow$)& 0.994 $\pm$ 0.003 ($\uparrow$)\\  
    
    \bottomrule    
  \end{tabular}

  \begin{tabular}{llllllll}
   \toprule
  \multicolumn{7}{c}{UNSW-NB15}\\
      \cmidrule(lr){2-7} 
   Label data ratio &  seen-AUT (B) & seen-AUT (A) & unseen-AUT (B) & unseen-AUT (A) & overall-AUT (B) & overall-AUT (A)\\
   \midrule
   
    2\% & 0.999 $\pm$ 0.000 & 0.460 $\pm$ 0.183 & 0.998 $\pm$ 0.000 & 0.570 $\pm$ 0.072 & 0.999 $\pm$ 0.000 & 0.539 $\pm$ 0.114\\
    10\% & 0.999 $\pm$ 0.000 & 0.769 $\pm$ 0.008 ($\uparrow$)& 0.998 $\pm$ 0.000 & 0.682 $\pm$ 0.009 ($\uparrow$)& 0.999 $\pm$ 0.000 & 0.729 $\pm$ 0.052 ($\uparrow$)\\           
     20\% & 0.999 $\pm$ 0.000 & 0.845 $\pm$ 0.012 ($\uparrow$)& 0.999 $\pm$ 0.000 ($\uparrow$)& 0.727 $\pm$ 0.018 ($\uparrow$)& 0.999 $\pm$ 0.000& 0.783 $\pm$ 0.014 ($\uparrow$)\\  
    
    \bottomrule    
  \end{tabular}

  \begin{tabular}{llllllll}
   \toprule
  \multicolumn{7}{c}{CICIDS-2017}\\
      \cmidrule(lr){2-7} 
   Label data ratio & seen-AUT (B) & seen-AUT (A) & unseen-AUT (B) & unseen-AUT (A) & overall-AUT (B) & overall-AUT (A)\\
   \midrule
   
    2\% & 0.999 $\pm$ 0.000 & 0.996 $\pm$ 0.001 & 0.996 $\pm$ 0.000 & 0.943 $\pm$ 0.006 & 0.998 $\pm$ 0.000 & 0.962 $\pm$ 0.006\\
   10\%& 0.999 $\pm$ 0.000 & 0.996 $\pm$ 0.001 & 0.999 $\pm$ 0.000 ($\uparrow$)& 0.990 $\pm$ 0.001 ($\uparrow$)& 0.999 $\pm$ 0.000 ($\uparrow$)& 0.993 $\pm$ 0.000 ($\uparrow$)\\
     
      20\% & 0.999 $\pm$ 0.000 & 0.995 $\pm$ 0.003 ($\downarrow$)& 0.998 $\pm$ 0.000 ($\downarrow$)& 0.982 $\pm$ 0.007 ($\downarrow$)& 0.999 $\pm$ 0.000 & 0.989 $\pm$ 0.003 ($\downarrow$)\\  
    
    \bottomrule    
  \end{tabular}

  \begin{tabular}{llllllll}
   \toprule
  \multicolumn{7}{c}{CSE-CICIDS-2018}\\
      \cmidrule(lr){2-7} 
   Label data ratio &  seen-AUT (B) & seen-AUT (A) & unseen-AUT (B) & unseen-AUT (A) & overall-AUT (B) & overall-AUT (A)\\
   \midrule
   
    2\% & 0.999 $\pm$ 0.000 & 0.857 $\pm$ 0.010 & 0.999 $\pm$ 0.000 & 0.348 $\pm$ 0.067 & 0.999 $\pm$ 0.000 & 0.544 $\pm$ 0.029\\
    10\% & 0.999 $\pm$ 0.000 & 0.832 $\pm$ 0.042 ($\downarrow$)& 0.999 $\pm$ 0.000 & 0.448 $\pm$ 0.126 ($\uparrow$)& 0.999 $\pm$ 0.000 & 0.577 $\pm$ 0.063 ($\uparrow$)\\
      
      20\% & 0.999 $\pm$ 0.009 & 0.761 $\pm$ 0.016 ($\downarrow$)& 0.999 $\pm$ 0.000 & 0.725 $\pm$ 0.100 ($\uparrow$)& 0.999 $\pm$ 0.000 & 0.675 $\pm$ 0.065 ($\uparrow$)\\  
    
    \bottomrule    
  \end{tabular}
  
 \end{table*}
\begin{figure*}[!htb]

 \centering 
\subfloat[\small{CTU-13}]{\label{ctu13} \includegraphics[scale=0.25]{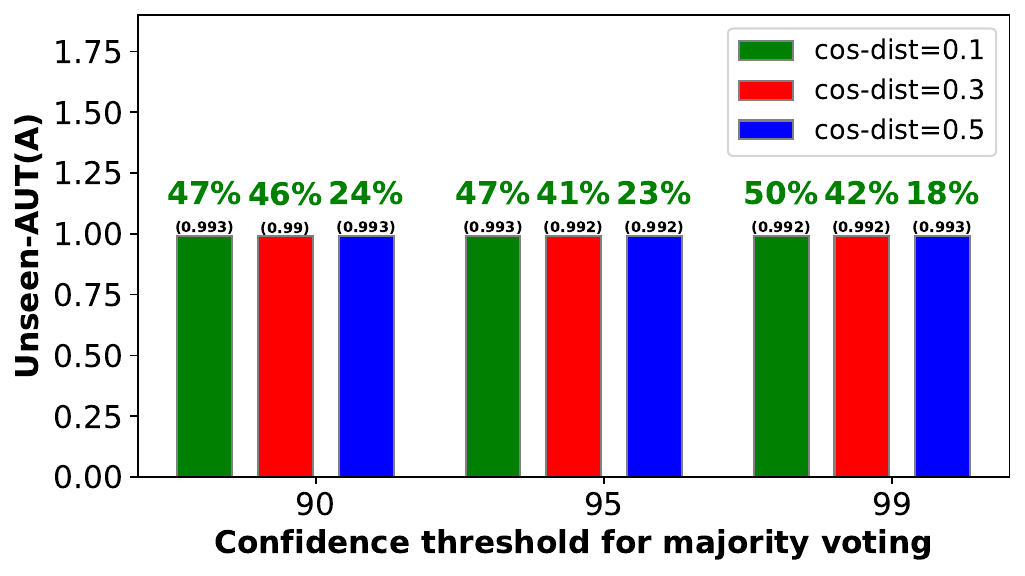}}
\subfloat[UNSW-NB15]{\label{unswnb15} \includegraphics[scale=0.25]{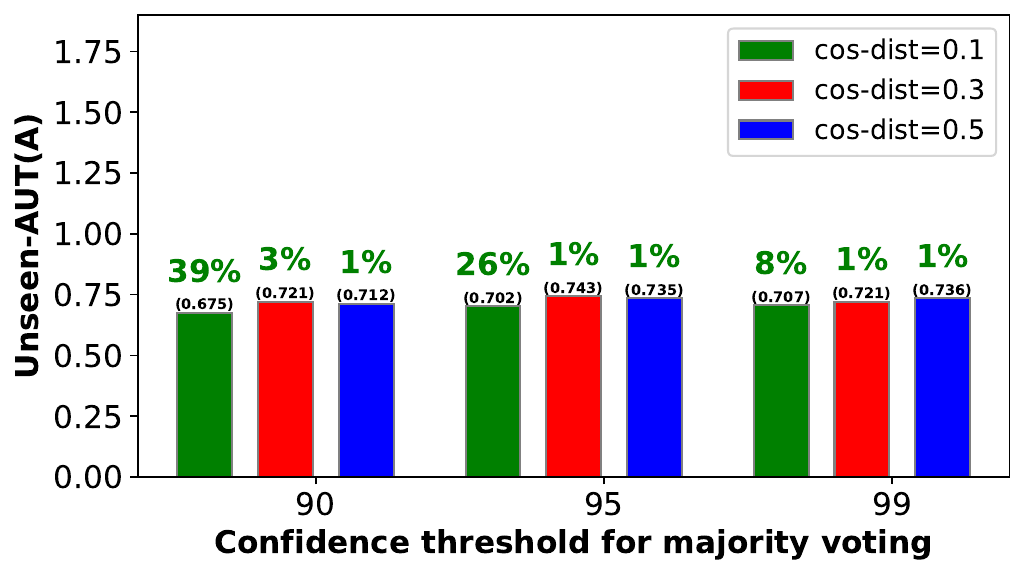}}
\subfloat[CICIDS-2017]{\label{ids17} \includegraphics[scale=0.25]{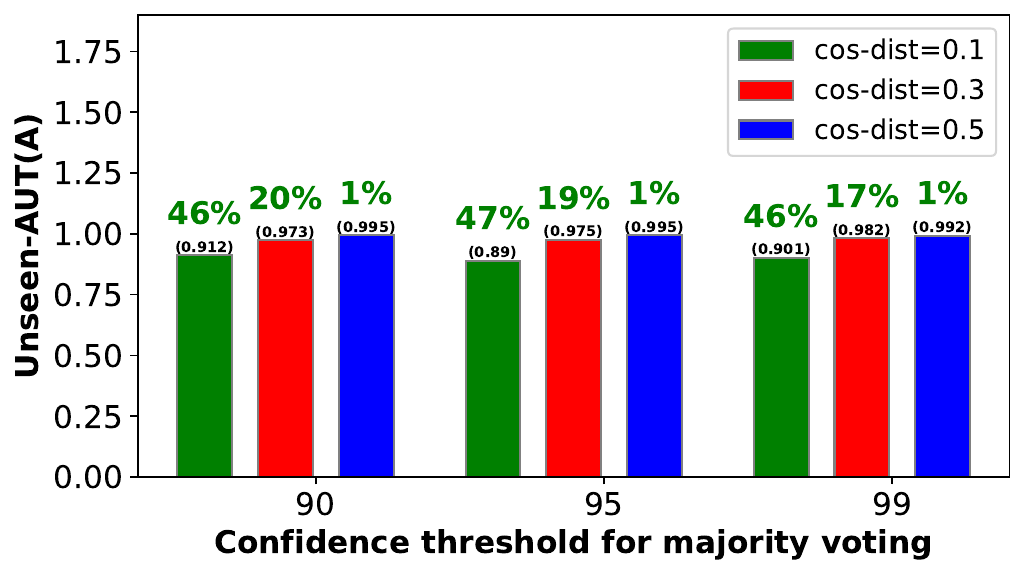}}%
 \subfloat[CSE-CICIDS-2018]{\label{ids18} \includegraphics[scale=0.25]{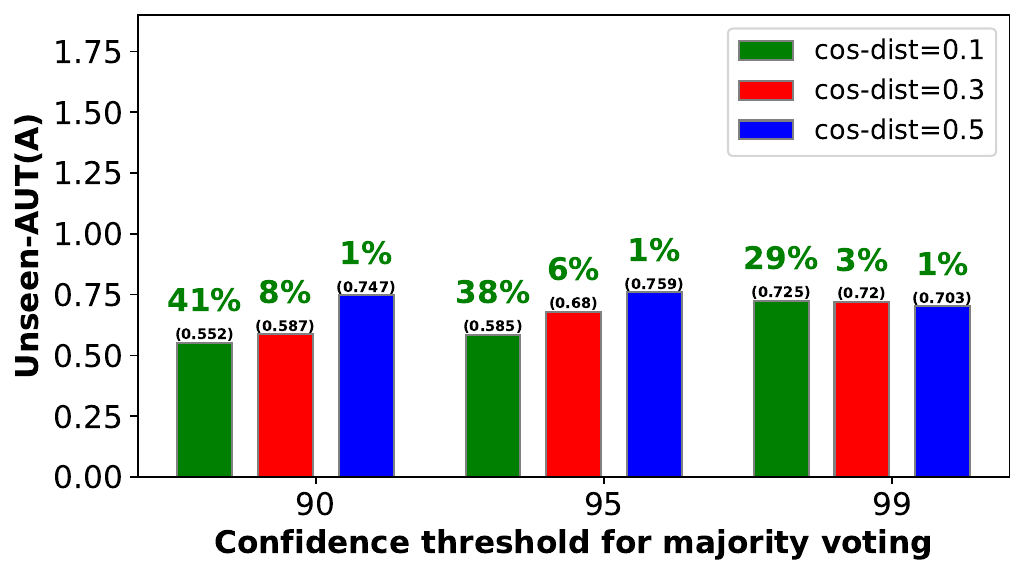}}%
\caption{Demonstrating the sensitivity of the cosine distance and majority voting scheme to the percentage of labeling effort saved and the detection performance of the unseen attack class (unseen-AUT (A)). The percentage of savings in the labeling effort is displayed above the respective bar plot and marked in green.}
\label{fig:cos-dist-conf-thresh}
\end{figure*} 

\par \textbf{Performance with varying label data:} We analyze how the performance results varies with different label data ratios. From Table~\ref{tab:ablation_study_varying_label_ratio}, with an increasing ratio of labeled data, the performance of the SOUL consistently increases on datasets CTU-13, UNSW-NB15, CICIDS-2017. On the CICIDS-2017 dataset when the labeled data ratio increased 20\% (from 10\%) seen AUT (A), unseen AUT(A), and overall AUT (A) reduced, however the difference in the third decimal with an average difference of 0.0021 so it impact is negligable.  We observe an exception, especially on CSE-CICIDS-2018 in which seen AUT (A) decreases with increasing labeled data ratio. On contrary, unseen AUT (A) and overall AUT (A) increases. Comparatively, this dataset that has nearly 63 million data samples and a higher class imbalance ratio per task than other datasets. This intuitively challenges the learning capabilities of the classifier. As a testament to our claim regarding learning capability, even the fully supervised baselines struggle to achieve better attack class detection on this dataset (refer to Table~\ref{tab:soul_comparison_with_unseen}). 
\par \textbf{Labeling effort savings with varying cosine distance and confidence threshold:} The percentage of savings on the labeling effort varies with cosine distance ($c_d$) and confidence threshold of the majority voting scheme, as demonstrated in Figure~\ref{fig:cos-dist-conf-thresh}. The larger cosine distance allows the noisy samples from memory to participate in the majority voting, which increases the variability in the label selection process. This behavior consistently affects the savings in labeling efforts across all the datasets. Furthermore, the confidence threshold during the majority voting also significantly affects the unseen-AUT (A) and labeling effort. For instance, with $0.1$ cosine distance the AUT (A) value changes to $0.725$ from $0.585$ when the threshold changes from $95\%$ to $99\%$ on CSE-CICIDS-2018 dataset. 

  \subsection{Details of the Hyperparameters}
\label{Section:hyperparameters}

\par \textbf{Hyperparamaters}: In our experiments, we employ the stochastic gradient descent (SGD) optimizer with a Nesterov momentum of 0.9. For all datasets, we use multistep LR with a decay factor of 0.96 at each epoch step and an early stopping strategy on the validation set, setting a patience value of 3 and a delta error threshold of 0.01. The train data is split into 71\% for training, 4\% as a validation set, and 25\% as a test set. The best hyperparameters are chosen using the validation set. For all experiments, we use fully connected (FC) multi-layer perceptron (MLP)s with 5-6 hidden layers with ReLU activations with batch normalization and dropout layer with probability of 0.2, followed by a softmax output layer of size two. The number of projection samples is 10,000 across all the experiments.  The details of the various hyperparameters are presented in Table~\ref{tab:hyperparameters}. The best hyperparameters for learning rate are chosen from the list [$10^{-1},10^{-2},10^{-3},10^{-4}$], weight decay from [$10^{-1},10^{-2},10^{-3},10^{-4}$], cosine distance from [$0.1,0.3,0.5$], and confidence threshold from [$90\%,95\%,99\%$] using the validation set of each dataset.

\begin{table}[htb]
  \caption{Hyperparamter details of the various benchmark datasets}
  \label{tab:hyperparameters}
  \centering
  \begin{adjustbox}{width=1\columnwidth}
  \begin{tabular}{llllllllllllllllllllllllllllll}    
     \cmidrule(lr){2-5} 
   Dataset & Batch size & Input size & Memory size & Architecture\\
   \midrule
    CTU-13 & 1024 & 39 & 1500 & FC:100,150,50,10,2\\ 
    UNSW-NB15 & 1024 & 202 & 6666 & FC:100,250,500,150,50,2\\
    CICIDS-2017 & 1024 & 84 & 13334 & FC:100,250,500,150,50,2\\
    CSE-CICIDS-2018 & 1024 & 84 & 13334 & FC:100,250,100,200,50,10,2\\
    \bottomrule
    \cmidrule(lr){2-5} 
   Dataset & Learning rate & Weight decay & Cosine distance & Confidence\\
   \midrule
    CTU-13 & $10^{-3}$ & $10^{-2}$ & 0.1 & 98\%\\ 
    UNSW-NB15 & $10^{-2}$  & $10^{-2}$& 0.1 & 98\%\\
    CICIDS-2017 & $10^{-2}$ & $10^{-3}$& 0.3 & 99\%\\
    CSE-CICIDS-2018 & $10^{-2}$  & $10^{-3}$ & 0.1 & 98\%\\
    \bottomrule
  \end{tabular}
  \end{adjustbox}
\end{table}
\par \textbf{Implementation details:} We utilize the open-source continual learning library Avalanche (version 0.2.1) for baseline methods such as EWC, SI, GEM and A-GEM. 
Additionally, we implement MIR, CBRS, and the proposed SOUL methods using the PyTorch library (version 1.13.0).  We conducted our experiments on a system with the following specs: $376$ GB of memory, $104$ cores (Intel(R) Xeon(R) Gold $6230$R CPU @ $2.10$GHz), and $2$ Nvidia Quadro RTX $5000$ GPUs. For reproducibility, we made the code and datasets available at this \href{https://github.com/amalapuram/SOUL}{link}.

\section{Summary}
\par 
The proposed method is motivated by our empirical observation that using gradient projection memory (constructed using buffer memory samples) can significantly improve the detection performance of the attack (minority) class when trained using partially labeled data. Further, using the classifier's confidence in conjunction with buffer memory, SOUL generates high-confidence labels whenever it encounters OWL tasks closer to seen tasks, thus acting as a label generator. Interestingly, SOUL efficiently utilizes samples in the buffer memory for sample replay to avoid catastrophic forgetting, construct the projection memory, and assist in generating labels for unseen tasks. 
\bibliography{References}
\bibliographystyle{ieeetr}
\vspace{-12mm}
\begin{IEEEbiographynophoto}{Suresh Kumar Amalapuram} recently completed his Ph.D. from computer science and engineering department at IIT Hyderabad. His research areas include continual learning, network intrusion detection, anomaly detection, and computer vision. Some of his earlier work was published in peer-reviewed conferences like IEEE INFOCOM, NeurIPS, COMSNETS, etc.
\end{IEEEbiographynophoto}
\vspace{-14mm}
\begin{IEEEbiographynophoto}{Shreya Kumar} recently completed her bachelors degree from computer science and engineering department at IIT Hyderabad. Currently, she is working as a software engineer at Arcesium India. Her research interests include machine learning, cybersecurity, and computer vision. 
\end{IEEEbiographynophoto}
\vspace{-14mm}
\begin{IEEEbiographynophoto}{Bheemarjuna Reddy Tamma}
is currently a Full professor in the computer science and engineering department at IIT Hyderabad. He received a Ph.D. from IIT Madras, and his research areas include converged cloud radio access networks, 5G, SDN/NFV, IoT/M2M, network security, wireless communication and networks, and cyber security. He published over 100 research articles in peer-reviewed conferences and journals, including IEEE INFOCOM, GLOBECOMM, NeurIPS, and IEEE Transactions on Green Communications and Networking.
\end{IEEEbiographynophoto}
\vspace{-14mm}
\begin{IEEEbiographynophoto}{Sumohana S Channappayya}
is currently a Full professor in the electrical engineering department at IIT Hyderabad. He received a Ph.D. from the University of Texas at Austin, and his research areas include Image and Video Quality Assessment, Biomedical Imaging, and Machine Learning. He published over 100 research articles in peer-reviewed conferences and journals, including ICIP, IEEE INFOCOM, NeurIPS, IEEE Transactions on Image Processing, and IEEE Signal Processing Letters.
\end{IEEEbiographynophoto}
\end{document}